# Materials Discovery in Combinatorial and High-throughput Synthesis and Processing: A New Frontier for SPM

Boris N. Slautin[1, a)], Yongtao Liu[2], Kamyar Barakati[3], Yu Liu[3], Reece Emery[3], Seungbum Hong[4], Astita Dubey[1,3], Vladimir V. Shvartsman[1], Doru C. Lupascu[1], Sheryl L. Sanchez[3], Mahshid Ahmadi[3], Yunseok Kim[5], Evgheni Strelcov[6], Keith A. Brown[7], Philip D. Rack[3], and Sergei V. Kalinin[3,8, b)]

[1] Institute for Materials Science and Center for Nanointegration Duisburg-Essen (CENIDE), University of Duisburg-Essen, Essen, 45141, Germany

[2] Center for Nanophase Materials Sciences, Oak Ridge National Laboratory, Oak Ridge, TN 37831, USA

[3] Department of Materials Science and Engineering, University of Tennessee, Knoxville, TN 37996, USA

[4] Department of Materials Science and Engineering, KAIST, Daejeon, 34141, Republic of Korea

[5] School of Advanced Materials Science and Engineering, Sungkyunkwan University (SKKU), Suwon 16419, Republic of Korea.

[6] Physical Measurement Laboratory, National Institute of Standards and Technology, Gaithersburg, MD 20899, USA

[7] Department of Mechanical Engineering, Boston University, 110 Cummington Mall, Boston, MA, 02215 USA

[8] Pacific Northwest National Laboratory, Richland, WA 99354, USA

[a), b)] Authors to whom correspondence should be addressed: boris.slautin@uni-due.de and sergei2@utk.edu

For over three decades, scanning probe microscopy (SPM) has been a key method for exploring material structures and functionalities at nanometer and often atomic scales in ambient, liquid, and vacuum environments. Historically, SPM applications have predominantly been downstream, with images and spectra serving as a qualitative source of data on the microstructure and properties of materials, and in rare cases of fundamental physical knowledge. However, the fast-growing developments in accelerated material synthesis via self-driving labs and established applications such as combinatorial spread libraries are poised to change this paradigm. Rapid synthesis demands matching capabilities to probe structure and functionalities of materials on small scales and with high throughput. SPM inherently meets these criteria, offering a rich and diverse array of data from a single measurement. Here, we overview SPM methods applicable to these emerging applications and emphasize their quantitativeness, focusing on piezoresponse force microscopy, electrochemical strain microscopy, conductive, and surface photovoltage measurements. We discuss the challenges and opportunities ahead, asserting that SPM will play a crucial role in closing the loop from material prediction and synthesis to characterization.

## I. Introduction

Materials discovery has been the bedrock of civilization, driving progress and transforming societies throughout history. The development initiated in the Stone, via Bronze to the Iron Ages. The invention of paper has had an enormous impact on the evolution of humanity serving as a pivotal medium for spreading and sharing knowledge since the creation of papyrus in ancient Egypt to the present day. Steel development revolutionized construction and manufacturing, enabling the creation of skyscrapers, railways, and countless tools. Porcelain, originating in ancient China, became a highly prized material, its discovery and refinement influencing global trade and cultural exchanges. Advances in energy materials have enabled the invention and optimization of batteries, a cornerstone of portable power, facilitating the rise of modern electronics and electric vehicles.[1, 2] Semiconductors, the material foundation of the information age, have undergone extensive research and development, leading to the sophisticated devices that define contemporary life.[3, 4] Similarly, the discovery of synthetic rubber and other polymers has revolutionized industries from packaging to aerospace, and the development of molecules for drugs has transformed medicine, offering cures and treatments for countless diseases.[5, 6] In each of these cases, a discovery was followed by multiple optimization cycles, often spanning decades or even centuries, highlighting the intricate and multifaceted process of materials innovation.

Historically, materials discovery has spanned a multitude of forms, ranging from serendipitous discoveries to knowledge-based optimization. Steel benefited from accidental innovations as well as systematic advances in alloy composition and processing techniques. The discovery of porcelain involved a blend of empirical experimentation and artistic innovation, eventually leading to highly specialized production methods. Batteries have evolved through both unexpected breakthroughs and targeted research efforts aimed at improving energy density and longevity. Semiconductors, initially discovered through chance, have since been optimized through rigorous scientific inquiry and advanced computational methods. Polymers were often discovered by accident, such as the development of nylon, while others were the result of targeted research, leading to a wide array of materials with diverse applications. Drug molecules have been discovered through both serendipity, like penicillin, and through methodical research and development processes, such as the development of targeted cancer therapies.

The development of machine learning (ML) offers a way to extend the paradigm of materials discovery. While earlier efforts are known, a significant inflection point occurred in 2006. At that time, Gerbrand Ceder's paper demonstrated the benefits of combining ML with modeling to the broader community.[7] Additionally, the launch of the Amazon Web Services (AWS) cloud platform, perhaps less well-realized at the time, provided scientists with the opportunity to run simulations from anywhere without the need to purchase, set up, or maintain dedicated clusters. This dual advancement illustrated the transformative potential of ML and cloud computing in scientific research.[8, 9] Much attention has been attracted to the acceleration of discovery via computation and ML, leading to exponential growth in computational efforts to predict new molecules and materials over the last 15 years.[10-13]

However, this progress necessitates experimental realization to validate and implement these predictions, ensuring that theoretical advancements translate into practical applications. On the synthesis side, the first approaches for accelerating materials synthesis via combinatorial libraries have been known since the late 1960s and have experienced several cycles of interest, intriguingly reminiscent of the AI winters.[14, 15] Automated synthesis has been broadly explored in industry, paving the way for advancements in materials discovery. Post-2010, several visionaries such as Lee Cronin, Benji Maryama, and others launched the first publicized synthesis efforts centered on developing automated synthesis platforms and labs.[16-21] The number of these research groups has been steadily growing, and since around 2020, the broad availability of commercial microfluidic and pipetting robots has given rise to multiple research groups in the U.S. and worldwide.[22] These technologies have enabled high-throughput synthesis, significantly accelerating the discovery and optimization of new materials.

In analyzing the experimental aspects, it becomes clear that accelerated materials discovery requires a transition to smaller length scales and shorter timescales. The necessity for smaller scales can be understood simply as a consequence of scaling throughput; synthesizing 10,000 compounds each by the gram is impractical. Instead, working at micro or nano scales allows for the rapid production and testing of numerous samples. Time is equally critical because rapid feedback is essential for efficient iteration and optimization. If information on the target functionality of a material is available within an hour or a day, it significantly accelerates the discovery process, whereas longer feedback times become a bottleneck. It is important to note that small-scale functionality does not always perfectly predict long-term macroscopic behavior. Nevertheless, small-scale measurements often serve as effective proxies for the desired properties, allowing researchers to efficiently screen and prioritize promising candidates for further development and rigorous validation at larger scales. This approach leverages high-throughput synthesis and characterization techniques to explore vast chemical spaces efficiently, thereby revolutionizing the speed and efficiency of materials discovery.

Here we pose that scanning probe microscopy (SPM) can become a key tool for closing the loop in rapid materials synthesis and processing. SPM offers high-resolution nanoscale insights necessary to evaluate the properties and functionalities of materials synthesized in combinatorial libraries.[23] However, to effectively leverage SPM in this capacity, it is crucial to integrate ML to handle the high throughput and large volumes of data generated. ML can match the speed and scale of SPM by automating the analysis of complex datasets, identifying patterns, and predicting material behaviors. This integration allows for real-time feedback and iterative optimization, effectively linking synthesis, characterization, and modeling. By incorporating SPM data into the materials discovery pipeline, researchers can rapidly screen and optimize new materials, ensuring that the vast amounts of information from combinatorial libraries are utilized efficiently. Ultimately, the synergy between SPM and ML will accelerate the discovery and development of materials with targeted functionalities pushing the boundaries of what is achievable in materials science as of now.

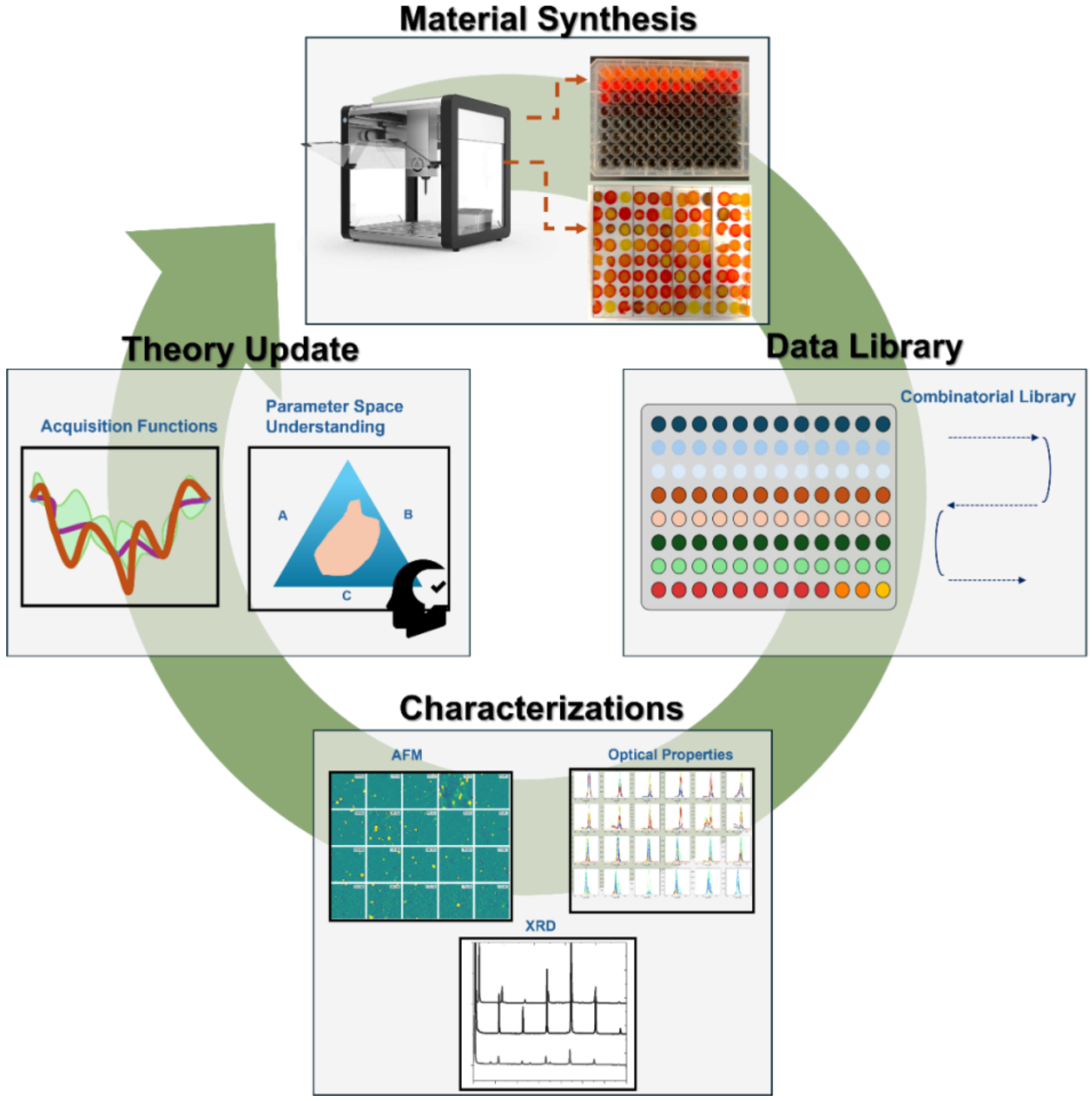

**Figure 1.** Schematic representation of the material discovery loop: theory → synthesis → characterization by multiple techniques → theory update based on the newly acquired data.

## II. Materials synthesis with rapid throughput

Acceleration of materials synthesis is a key requirement for the overall acceleration of the materials discovery cycle. Unlike the theoretical workflows that can be scaled almost agnostically of the specific theory type, synthesis methods are highly specific and are associated with their own balance on latencies and multiplexity. For example, microfluidic systems allow for rapid sequential fabrication and in principle rapid feedback, whereas combinatorial methods generate spread libraries containing an exhaustive sampling of chosen composition space, but the feedback and transition to a different composition or processing space are associated with significant latencies. These considerations, in turn, inform the development of characterization workflows. Here, we provide a brief overview of the synthesis methods as a necessary upstream task for characterization.

### II.a. Combinatorial spread libraries

Combinatorically deposited films can be leveraged as a rapid materials discovery and optimization process in a variety of material and application spaces. This is done through the synthesis of a wide targeted composition range in conjunction with high throughput screening methods to enhance understanding of composition-structure-property relationships to identify compositions with desired properties.[24, 25] This approach was first developed in the 1960's as a way to explore isothermal ternary phase diagram spaces[26] and was further leveraged to search for oxide superconducting materials.[27] Since then, this technique has been employed for a variety of material discovery processes. One of the primary use cases for the combinatorial approach is in the development of functional materials such as the superconducting materials mentioned above. Other work in this area includes transducer materials,[28] hydrogen storage,[29] fuel cell electrodes,[30, 31] and lithium-ion battery electrodes.[32] Another interesting use case of thin films for rapid alloy assessment is for optical data storage utilizing phase change materials.[33, 34] In the post-COVID world, anti-microbial material properties have also been of interest. The rapid solidification into metastable materials and the discovery process utilizing thin films can be used to investigate these properties[35] to limit transmission of pathogens in high-traffic touch surfaces such as doorknobs, countertops, and handrails among others. The combinatorial thin film approach also allows for the sustainable development of new alloys. Traditional bulk metallurgy techniques have high time, energy, and material demands. However, the combinatorial approach reduces all three of these demands allowing for a faster, more sustainable methodology.[36] Once composition ranges of interest are identified, more conventional methodologies can be leveraged to confirm the bulk structure-processing-properties relationship.

A variety of synthesis techniques exist and can be broken down into two main categories: physical (PVD) and chemical vapor deposition (CVD). Combinatorial chemical vapor deposition techniques include spatially separated reactant deposition[37-40] and chemical beam vapor deposition.[41, 42] CVD is less prevalently used but is leveraged in microelectronics where the non-line-of-sight process is an advantage. PVD being the more common methodology includes techniques such as evaporation,[43] pulsed laser deposition, [44, 45] molecular beam epitaxy, [46] and sputtering.[47] Combinatorial sputtering is the most widely leveraged technique as a convenient tool for rapid material discovery, because it can achieve thin film compositions that cover a large and tunable composition range.[48-51] Higher complexity combinatorial films can be attained through additional sputtering targets and power supplies. Appropriate configuration of the additional sputtering targets can create a wide compositional gradient across a single wafer. Alternatively, the substrate can be rotated to generate a film of uniform composition and thickness. Further, the high energy deposition process and fast cooling rates of the energetic sputtered species allow for the realization of metastable material configurations, for example, supersaturated solid solution found in various as-deposited sputtered alloys.[49, 50] Subsequent annealing can be utilized to achieve the equilibrium state of the system through recrystallization, grain growth, and phase separation.[51, 52]

Combinatorically synthesized films have also been utilized in mechanical alloy design for mechanical, thermal, and corrosion properties as well as microstructure determination. It has proven to be a powerful tool for understanding the composition-structure-property relationship and identifying composition ranges of interest for enhanced or unique material properties.[53] Previously, Al-Si alloys have been investigated using the combinatorial thin film technique. Al-Si alloys are of interest especially in the automotive industry due to their high strength-to-weight ratio, thermal conductivity, and low thermal expansion.[54] This system has thus been further investigated to understand the role of Si content on the mechanical properties as well as influence on microstructure. Olk *et. al.* investigated Si contents from approximately 15 to 85 at. % Si combinatorically and showed a strong dependence of phase formation and hardness on the Si content.[55] Further work by this group investigated the microstructure and the ability to tune it in the Al-Si system. They showed that an amorphous phase *a*-(Al–Si) and a mixed phase a-Si and c-Al containing a tunable Al grain structure and engineered sample surface can be produced.[56] This work was done in parallel with nanoindentation to correlate the mechanical properties. More recently, combinatorial exploration has been performed in the Al alloy space leveraging Ce additions in search of low coefficient of thermal expansion (CTE) phases to replace the traditional Al-Si piston alloys mentioned previously.[57] Combinatorial nanoindentation work has also been performed in the Ti-Al system. One study explored this system from about 0 at. % to 35 at. % Al with near uniform thickness of around 900 nm and demonstrated that thin films can be leveraged to track mechanical property changes over a composition gradient using nanoindentation.[58] Further studies have been performed to understand the interaction effects between the film and substrate for mechanical property determination. As expected, the interaction is dominated by the thickness of the deposited film. However, parameters can be adjusted and thickness-corrections applied to accommodate the substrate interaction zone with previous work showing good agreement between thin film and upscaled materials properties such as the CTE and other mechanical properties.[57, 59-64] Additionally, corrosion behavior can be effectively investigated utilizing the combinatorial thin films rapid material discovery technique.[65] This approach has been applied to a variety of other alloy classes. High entropy alloys (HEAs) or compositionally complex alloys (CCAs) have been investigated using this technique as the optimization and development of these alloys can prove difficult and time-consuming using conventional metallurgical techniques.[66] Other related work with HEA thin films suggests that combinatorial thin films are a promising and cost-effective way to not only investigate properties (such as mechanical properties and corrosion behavior), but are also promising as tools for advanced coating materials.[67-69]

Combinatorial synthesis allows for rapid access to vast composition spaces across a variety of material application spaces including electronics, functional materials, and structural materials with the ultimate goal of correlating composition with desired properties. A particular emphasis has been placed on optical and magneto-plasmonic materials.[49-51, 70-76] Other functional materials such as catalytic,[77, 78] magnetic,[79, 80] and electrical[81] materials have also been explored. Structural materials have been explored with a focus on mechanical alloys[57, 59, 65, 82-84] and radiation hard materials.[85, 86] This then shifts the bottleneck from synthesis to characterization of functionality and other material properties. As such, it is then vital to develop a technique to increase throughput while maintaining optimized material functionality.

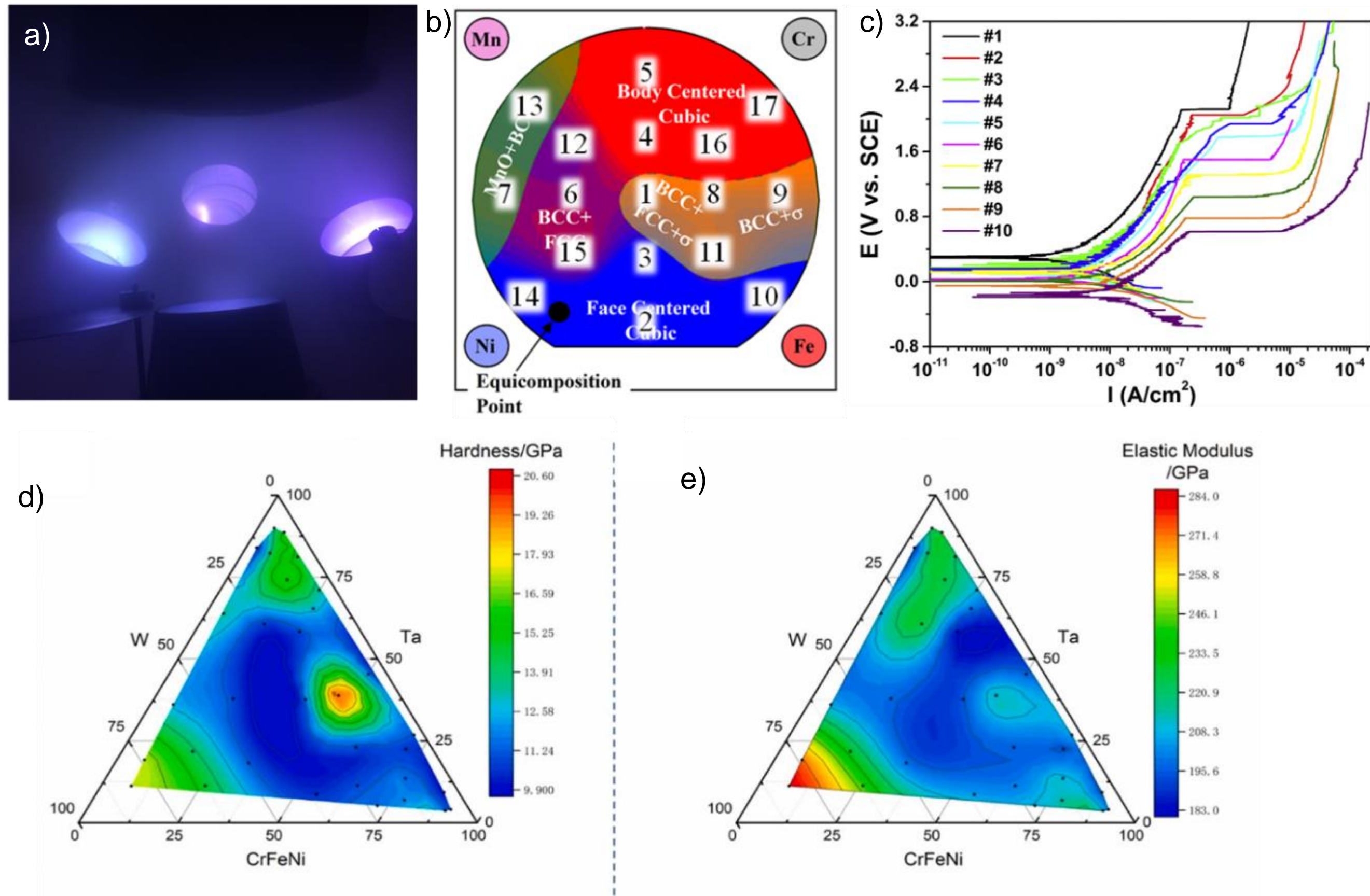

**Figure 2**. a) Digital photograph of a four target sputtering system in operation, b) Cr-Fe-Ni-Mn combinatorial wafer mapping crystal structure versus composition,[87] c) Tafel plots from an $Al_x(CoCrFeNi)_{1-x}$ combinatorial thin film versus composition. d) and e) show modulus and hardness mapped out through nanoindentation of a combinatorial film versus composition. c) Reprinted from [65] Copyright (2020), with permission from Elsevier. d) and e) Reprinted from [88] Copyright (2022), with permission from Elsevier.

### II.b. Laboratory robotics

Laboratory robotics is central in transforming materials discovery and optimization by enabling the automation of high-throughput synthesis and screening. Liquid handlers and mobile robotic systems, among other robots, have already played key roles in enabling fast experimentation with the least human intervention to provide precision and reproducibility. An increasing number of research laboratories are adopting such technologies that help speed up their workflow efficiently to explore wide chemical spaces.

In recent years, the integration of laboratory robotics has emerged as a transformative approach to overcome these challenges.[89-95] These systems, including automated pipetting robots, enable high-throughput synthesis and characterization, significantly reducing human error and ensuring high reproducibility. Additionally, they play a crucial role in accelerating the discovery, design, and optimization processes in materials science allowing researchers to explore vast parameter spaces more efficiently and effectively.[96-99]

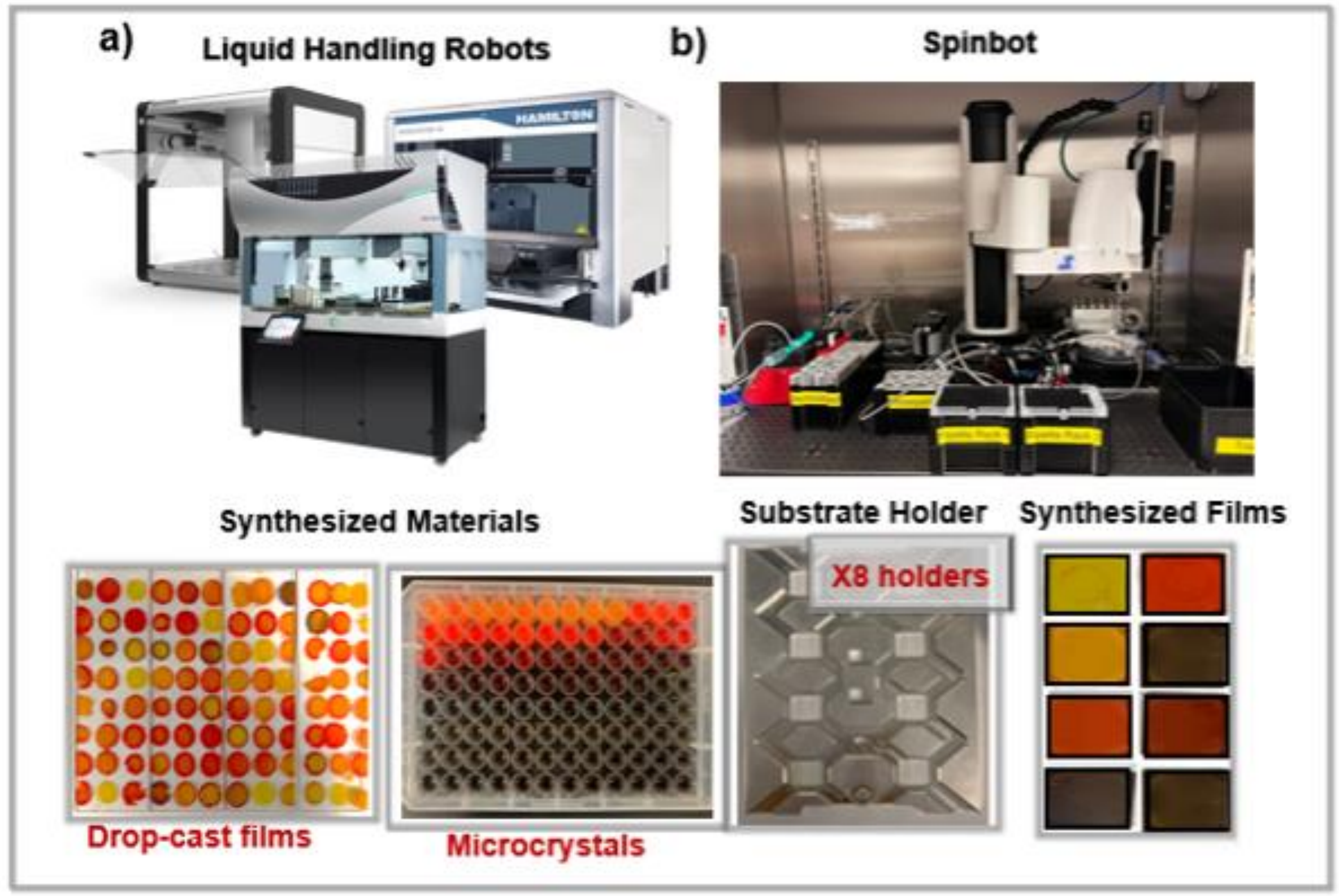

**Figure 3.** Examples of robotic platforms (a) liquid handling robots (b) spinbot for material research

In the automated settings, robots prepare and dispense precise reagent quantities, mix solutions, and carry out complex synthesis procedures as seen in Ahmadi et al.[97-100] , Brabec et al.[101-103] for solution-based processing of photovoltaic materials, and more.[104-107] Researchers in Ahmadi's group have utilized advanced robotic systems to synthesize a variety of halide perovskite materials, including 3D, 2D, quasi-2D, and nanocrystals for optimization and explorations of different phenomena in this complex materials system. Similarly, Langner et al.[102] used liquid handling robotics to formulate inks for organic photovoltaic film fabrication. These works highlight the efficiency and scalability of liquid-handling robots in material synthesis.

In addition to automated synthesis, vast data can be generated via high-throughput characterization methods such as time-dependent photoluminescence properties of a combinatorial library of optoelectronic materials like halide perovskites to study the phase evolution and stability in these materials systems. This technique allows for real-time monitoring of the materials properties, providing insights into the dynamics of phase evolution, kinetics of growth and stability. Extensive phase diagrams, such as ternary diagrams, are constructed to investigate the interplay between various dimensions of materials and their impact on stability, optical band gaps, and phase changes.

Automating these processes ensures high reproducibility, reduces human error, and minimizes material waste. This efficiency is achieved, because automated systems are designed to handle very small volumes of materials, enabling precise investigations with minimal consumption. Some of these highly accurate robotic systems can work with volumes smaller than 0.1 microliters,[108] which provides an overview of their capabilities and precision. By conserving valuable materials and ensuring exceptional accuracy, these systems empower researchers to efficiently explore complex parameter spaces while maintaining strict control over experimental conditions. Solution processing of materials usually promises low-cost and large-scale manufacturability. However, the final products of these materials can be affected by various synthesis parameters, such as solvent, antisolvent, concentrations, temperature, and ambient conditions, all of which can be controlled and explored using synthesis robots as well. This advancement is particularly beneficial for the synthesis of semiconductor materials like perovskites, where precise control over composition and processing conditions is essential for achieving the desired functionalities. These materials are especially challenging because of their vast compositional and structural spaces, which significantly limits the feasibility of manual exploration. Automated systems address this challenge by enabling high-throughput investigations that efficiently navigate these complex spaces.

Mechanical robotic fabrication is especially important for van der Waals solids[92, 95] that up to date have not been synthesized by any other means. The developed systems allow for automated identification of single-layer 2D material flakes that were mechanically exfoliated onto a substrate with subsequent robotic transfer stacking of the identified flakes with the needed twist angle precision. This mechanical synthesis process can produce materials not existing in nature comprising arbitrary stacking sequences of dissimilar 2D chemical composition lattices for engineering the optical, magnetic and electrical properties of the materials.

The development of automated laboratories has advanced to include self-driving laboratories (SDLs), [107, 109, 110] which leverage technologies such as automated thin-film deposition robots (often referred to as "spinbots") [111-113] , and free-roaming robotic systems.

[114, 115] These innovations aim to shift automation from individual instruments to broader research workflows, effectively creating what can be described as "robotic chemists."

Utilizing a robot to perform repetitive and time-consuming tasks, especially for toxic materials, is beneficial but many complex tasks still rely on human intervention and expertise such as decision-making, handling of materials, and adjustment of the experimental set-up.

Efforts have also been directed towards developing self-driving labs, which integrate robotics with machine learning algorithms to enable autonomous decision-making during experiments. By continuously learning from experimental data, these systems can optimize synthesis protocols with minimal human intervention, paving the way for more efficient and scalable material discovery processes. A notable example is the A-Lab, developed by Zeng and Ceder at Berkeley's National Laboratory.[116] This lab represents a fully automated laboratory setup where no human intervention is required during the synthesis and characterization processes. The A-Lab identifies air-stable, unreported targets using density functional theory (DFT)-calculated convex hulls, consisting of ground states from the Materials Project and Google DeepMind. Synthesis recipes for each target are proposed using ML models trained on synthesis data from literature.

The robotic laboratory automates several critical steps: precise powder dosing, controlled sample heating, and product characterization using X-ray diffraction (XRD). All sample transfers between these stations are performed using robotic arms, forming a fully automated sequence from chemical input to characterization. Phase purity is assessed using XRD, analyzed by ML models trained on structures from the Materials Project and the Inorganic Crystal Structure Database (ICSD), and confirmed with automated Rietveld refinement. If high (>50 %) target yield is not obtained, new synthesis recipes are proposed by an active-learning algorithm that identifies reaction pathways with maximal driving force to form the target. The A-Lab has demonstrated the ability to synthesize 41 materials within 17 days, showcasing the efficiency and scalability of fully automated systems combined with machine learning decision-making models.

Laboratory robotics is transforming materials discovery and optimization by providing efficient, precise, and reliable alternatives to manual methods. Systems like liquid-handling robots and fully autonomous platforms such as A-Lab enable rapid high-throughput synthesis and data-driven optimization, allowing researchers to explore vast chemical spaces quickly. These technologies have demonstrated their value in tasks such as synthesizing thousands of microcrystals in days, automating van der Waals solid assembly, and scaling thin-film characterization. While challenges remain in system standardization and integrating human expertise, continued advancements in robotics are poised to accelerate material discoveries and innovations.

### II.c. Microfluidics

Other automated robotic platforms that have gained popularity are microfluidics[19, 117-120], which manipulate fluids at the microscale and have emerged as powerful tools for synthesizing new materials. Microfluidic devices allow precise control over reaction conditions, including mixing reactants, temperature, and flow rates within tiny channels. This control is beneficial for synthesizing materials with uniform properties and creating gradient libraries with systematic composition variations. Microfluidic systems, like those developed by Abolhasani et al.[121, 122], who developed the artificial chemist and smart dope as a self-driving fluidic lab, are able to accelerate synthesis space exploration and optimization of lead halide perovskite quantum dots (LHPs) (QDs). DeMello et al.[123] were able to successfully synthesize Pt and Pd single atom stabilized on graphitic carbon nitride utilizing a microfluidic system. Utilizing this microfluidic system, they conducted a systematic study on the effect of the metal on the final loading of single atom catalysis.

Microfluidic systems can also be incorporated with ML methods to create rapid, high-throughput screenings of various applications, even like drug combinations, as seen by Zhou and Huang et al.[124], where they incorporated artificial intelligence (AI) with microfluidics to create an AI-accelerated high-throughput combinatorial drug evaluation system to rapidly synthesize and screen drug combinations using a mask region-based convolutional neural network (Mask R-CNN) with ResNet-50, which is a library that provides image detection and segmentation algorithms to analyze the obtained images. Wang et al.[125] were able to utilize ML-assisted synthesis of perovskite quantum dots to elucidate the nucleation growth-ripening mechanisms. They were also able to guide the synthesis of the QDs with precise wavelength conditions by combing gradient boosting (GB) and random forest (RF) to make predictions of the photoluminescence (PL).

A key advantage of microfluidic systems is to be able to conduct multiple reactions in parallel on a single chip. Microchannels, individually controlled, perform distinct reactions simultaneously, significantly increasing material synthesis throughput. Microfluidic platforms integrate with analytical techniques like real-time spectroscopic monitoring, providing immediate feedback on reaction progress and outcomes. This integration enables rapid identification of promising material candidates and on-the-fly adjustments to synthesis parameters. The ability to conduct iterative optimization in real-time is crucial for refining material properties efficiently.

Microfluidics complement robotics by enabling precise control over microscale reactions and facilitating high-throughput experimentation. This technology has been instrumental in applications such as quantum dot synthesis and drug screening. Platforms like the "Artificial Chemist" demonstrate how real-time optimization in microfluidic systems can significantly accelerate material discovery. The ability to perform parallel reactions and integrate real-time analytical feedback allows researchers to efficiently explore complex chemical spaces with unparalleled precision. Note that microfluidics can be directly extended for rapid synthesis of the combinatorial and positionally encoded droplet libraries.

## III. SPM based fabrication

The potential for scanning probes to control matter at the highest resolution was dramatically shown in 1990 when Eigler and Schweizer positioned single xenon atoms using a scanning tunneling microscope.[126] This initial demonstration captured the imagination of the nanoscience community that sought advanced methods for lithography – or the process of defining patterns with particular relevance to the nanoelectronics industry.[127, 128] Over the years, it became clear that scanning probes can not only define high resolution patterns, but also control the composition, phase, and chemistry of materials with a unique combination of resolution and throughput. Further, the sensing features of scanning probe microscopes provide an ability to probe and characterize samples at their native scale at which they are made which enables immediate feedback and can thus accelerate the discovery of new materials.[129] In the following paragraphs, the most widely studied methods of SPM-based materials synthesis and preparation are summarized according to the type of material deposition or transformation.

Perhaps the first scanning probe method to perform meaningful material transformations at the nanoscale was oxidative scanning probe lithography. In this approach, the probe is used as one electrode of an electrochemical cell to electrochemically modify the sample at the tip location. Initially demonstrated as a method of chemically modifying the surface of hydrogen-terminated silicon,[130] this method has been extended to many other materials and types of interactions.[131] A common application of this method is the local conversion of metals, semiconductors, or dielectrics into an oxide through anodic oxidation. Standout examples of this process include silicon,[132] metals such as niobium,[133] diamond,[134] and more recently two-dimensional materials such as graphene[135] and tungsten diselenide.[136] While local oxidation is a fairly constrained material transformation, it has nonetheless been used to study material properties. For example, anodic oxidation has been used to produce quantum dots in graphene that can be electrically contacted to measure their transport properties.[137] Additionally, it was found that charged probes can write conductive patterns in the interface between thin $LaTiO_3$ films on $SrTiO_3$ substrates,[138] which was later attributed to a reversible electrochemical process controlling the surface hydroxylation,[139] and has subsequently been used to study quantum phenomena such as Kronig-Penney superlatices.[140] Furthermore, scanning probe-based patterning of phosphorus on the surface of hydrogenated silicon allowed for the realization of a single-atom transistor.[141] As shown by these examples, most studies of material properties enabled by oxidative scanning probe lithography have focused on electronic properties as these are strongly affected by the oxidation state of the material.

In addition to changes in the oxidative state, scanning probe microscopes have been used to locally change the phase of materials through a variety of means. For instance, thermal scanning probe lithography, or local heating using a scanning probe, can drive materials through a local phase transformation.[142] Such thermally driven phase changes have been extensively studied in the context of information storage. For example, $Ge_2Sb_2Te_5$ can be driven between amorphous and crystalline phases by modulating the thermal profile using a local scanning probe at the level of a terabit per square inch.[143, 144] Such processes have also been used to modulate optical properties to realize and study photonic metasurfaces.[145] Electric fields can be used to drive local phase changes in materials. Electrostatic force microscopy is the category of scanning probe microscopy associated with strong localized electric fields. This has been used in a number of contexts for manipulating material phases. For instance, vanadium oxide may be driven through its insulating-to-metal transition using local electric fields.[146] Further, topotactic phase transformations may be driven by local electric fields, such as the transformation between perovskite and brownmillerite $SrFeO_x$.[147] Such examples show how local stimuli, administered by scanning probes, can be used to modulate material phase and therefore study an important facet of processing-property relationships in materials.

In addition to modulating the phase of materials, scanning probes can also be used to directly deposit materials to realize nanoscopic samples for further characterization. Dip-pen nanolithography (DPN) was the first technology to demonstrate this process wherein a scanning probe tip was coated with a small molecule and then used to locally pattern this material into a self-assembled monolayer.[148] This first example took advantage of the water meniscus that forms during tip-sample contact in a humid environment and, as such, it led to the exploration of a wide array of water-soluble materials including polymers, proteins, and nanoparticles.[149-151] Subsequent work focused on removing the water solubility restriction and employed a variety of approaches. One method is to use a non-volatile solvent to transport materials of interest.[152, 153] Alternatively, heating the probe may melt nominally solid polymers to directly pattern them in an approach known as thermal DPN.[154] Finally, hygroscopic polymers can be used as a matrix to pattern a variety of materials that are not themselves soluble.[155] A particularly noteworthy approach to studying nanomaterials is scanning probe block copolymer lithography (SPBCL), in which nanoparticle precursor materials are transported in a polymer matrix and then subsequently processed to form spatially-encoded nanoparticles.[156, 157] This approach has been extended to synthesize metallic alloy nanoparticles made of up to five different metals.[158, 159]

In considering the vanguard of DPN and related direct transport techniques, there are two main approaches for integrating materials synthesis into a broader materials discovery pipeline. The first approach is considered the library approach in which the scanning probe system is used to construct large libraries of materials that can be selected individually for subsequent analysis. In this approach, throughput and material diversity are crucial considerations, as the goal is to produce as many materials as possible. The serial writing nature of scanning probes can be rate limiting and this challenge has motivated the development of parallel approaches including the development of multi-cantilever arrays.[160, 161] In an advance that lowered the complexity and cost of probe arrays, cantilever-free probe arrays produced using soft lithography have been developed that allow millions of probes to be operated in parallel in a low-cost format.[162] This approach can be combined with spray coating to produce gradients of composition[163] and gradients in applied force through intentional tilting can produce gradients in deposited material amount.[164] Nanoparticles made using these approaches have been screened for their optical [165, 166] and catalytic properties.[159, 163, 167]

In contrast with the library approach, the closed-loop approach is to make individual samples and then test their properties as they are prepared so that the extracted information can be leveraged to select the next sample.[129] While DPN is traditionally an open-loop process, there have been steps towards realizing closed-loop DPN. Specifically, it has been recognized that patterning can be detected in real time by monitoring the vibrational resonance of the cantilever,[168] and that this signal can be used as a feedback signal to obtain high precision and accuracy patterning[169] with samples at the femtogram scale.[170] There is even potential to combine the closed-loop and the library approaches as cantilever-free probe arrays can generate feedback through optical or electrical means[171, 172] and methods have been developed to independently actuate individual cantilever-free probes.[173, 174]

Collectively, scanning probes have shown a remarkable versatility in defining and transforming materials on the nanoscale. There are already promising examples of their fruitful application for materials discovery and as the techniques become more sophisticated, these examples are expected to continue to grow.[175]

## IV. SPM characterization

There are many scanning probe microscopy (SPM) methods, each with its unique capabilities and applications.[176] These include topography imaging, which maps the surface structure of materials at the nanoscale; phase imaging, which provides contrast based on material composition and mechanical properties; electrical methods, such as conductive atomic force microscopy (CAFM),[177, 178] which measure electrical conductivity and other electronic properties; and electromechanical techniques, like piezoresponse force microscopy (PFM),[179] which probe the electromechanical behavior of materials. Numerous reviews have extensively covered these diverse SPM methods, highlighting their principles and applications.[176, 178-184] In this context, we aim to formulate general requirements for SPM methods that can be integrated into self-driving labs for rapid materials characterization, both for the combinatorial library characterization and downstream from the fast synthesis. These requirements include high spatial resolution, quantitative measurement capabilities, compatibility with automated control, and robust data analysis frameworks, particularly incorporating machine learning (ML) techniques. Additionally, we provide specific examples of SPM methods that meet these criteria, illustrating their potential to enhance the efficiency and accuracy of material discovery processes.

A general requirement for SPM methods in the context of rapid material characterization is that the measurements must be sensitive to the core functionality of the material and can serve as either a direct measure or a reliable proxy for this functionality. Additionally, it is essential to minimize the influence of exogenous factors, such as topography, on the measurements. While topography itself can provide valuable information, it should be manageable and not interfere with the core functionality assessment.

### IV.a. Piezoresponse Force Microscopy

The visualization of domain structures on the surface of ferroelectric materials using electromechanical detection in SPM was first demonstrated in 1992.[185] This technique now known as Piezoresponse Force Microscopy (PFM), rapidly gained widespread use in the ferroelectric community.[186] Due to its non-destructive nature, low cost, and minimal sample preparation requirements – aside from the need for a reasonably smooth surface – it has become the gold standard for investigating local ferroelectric and piezoelectric properties, as well as ferroelectric domain structures.[182, 187-189]

PFM is based on measuring the local strain induced by the converse piezoelectric effect when an AC voltage is applied through the SPM probe in contact with the surface of the material (Figure 4a,b). The use of lock-in selective registration techniques for measuring the piezoresponse enables a sensitivity to surface displacements in the vertical direction of better than 1 pm.[190] The development of resonance tracking PFM methods, such as Dual Resonance Amplitude Tracking (DART)[191] and Band-Excitation PFM (BE PFM),[192] significantly improved the signal-to-noise ratio in PFM measurements by utilizing resonance enhancement. The spatial resolution of PFM is determined by the contact tip radius and can reach a few nanometers. The above discussion also applies to measure of lateral (in-plane) local surface displacements, which occur when there is a non-zero in-plane component of the polarization in the ferroelectric domain beneath the tip.[193] Beyond the routine visualization of domain structures, PFM offers a wide range of spectroscopic techniques, most of which were developed in the 1990s and early 2000s.[194-199] Analogous to macroscopic characterization, the most used spectroscopical technique are local electromechanical hysteresis loop measurements through cyclic switching with a modified triangular-shaped voltage waveform (Figure 4c), first demonstrated by Hidaka et al. in 1996.[200] This method offers deep insights into the local ferroelectric state and polarization switching dynamics including domain nucleation,[201] the influence of domain walls[202] and isolated defects on ferroelectric switching,[203] the specifics of switching across grain boundaries,[204] etc.

Despite the unique capabilities of PFM, several fundamental challenges complicate the interpretation of the measured signal and hinder quantitative analysis. The primary issue is the difficulty in isolating the 'pure' piezoresponse from the raw measured signal, which is often affected by electrostatic, electrochemical, and other non-ferroelectric sources.[205-210] Electrostatic influences during PFM measurements also involves charge injection beneath the probe. The accumulation of charges near the tip-sample contact can result in hysteresis-like behavior during loop measurements, masking the true ferroelectric response and leading to potential misinterpretation of experimental results.[211, 212] Additionally, this charge accumulation affects the polarization switching process itself, often complicating the creation of local domain structures with precise parameters using SPM.[213] Another challenge in PFM is the difficulty of quantitatively characterizing the measured parameters. Although the theory of PFM signal formation is well-developed,[80, 214-221] accurately determining ferroelectric

properties—particularly piezoelectric coefficients – requires accounting for a wide range of factors that are often difficult to measure. These include surface layer properties, electric field distribution, and various parasitic effects, such as those mentioned above. As a result, despite significant efforts and visible progress,[220, 222-225] achieving reliable quantitative PFM measurements of real materials remains largely an unresolved issue.

Since its invention, PFM has been widely used to characterize ferroelectric-related properties across a vast array of materials. The ability to manipulate and create isolated domains using a biased SPM tip sparked a wave of extensive PFM research into the properties of ferroelectric domains in various materials, including films, crystals, ceramics, and complex heterostructures.[187, 188, 226-228] The technique has provided invaluable insights into domain dynamics, domain wall motion, and switching mechanisms across these material classes.[179, 182] In particular, PFM has been crucial for studying the evolution of polarization states in relaxor ferroelectrics, where it has helped to uncover the complex mechanisms behind polarization relaxation and stability of poled states.[229-231] Combining lateral and vertical PFM with sample rotation in the XY plane (Vector PFM or 3D-PFM) has enabled the reconstruction of full polarization orientation patterns in three dimensions (Figure 4f).[232] Vector PFM has provided a more comprehensive understanding of the spatial distribution of polarization in materials, especially in ferroelectric ceramics and other complex systems, enabling the study of polarization textures, vortex-like domain structures, and topological defects in detail.[232-238] Beyond traditional ferroelectric systems, PFM has also been applied to investigate emerging materials like multiferroics, 2D materials,[239] and biological materials exhibiting piezoelectric or ferroelectric-like behavior.[240-243]

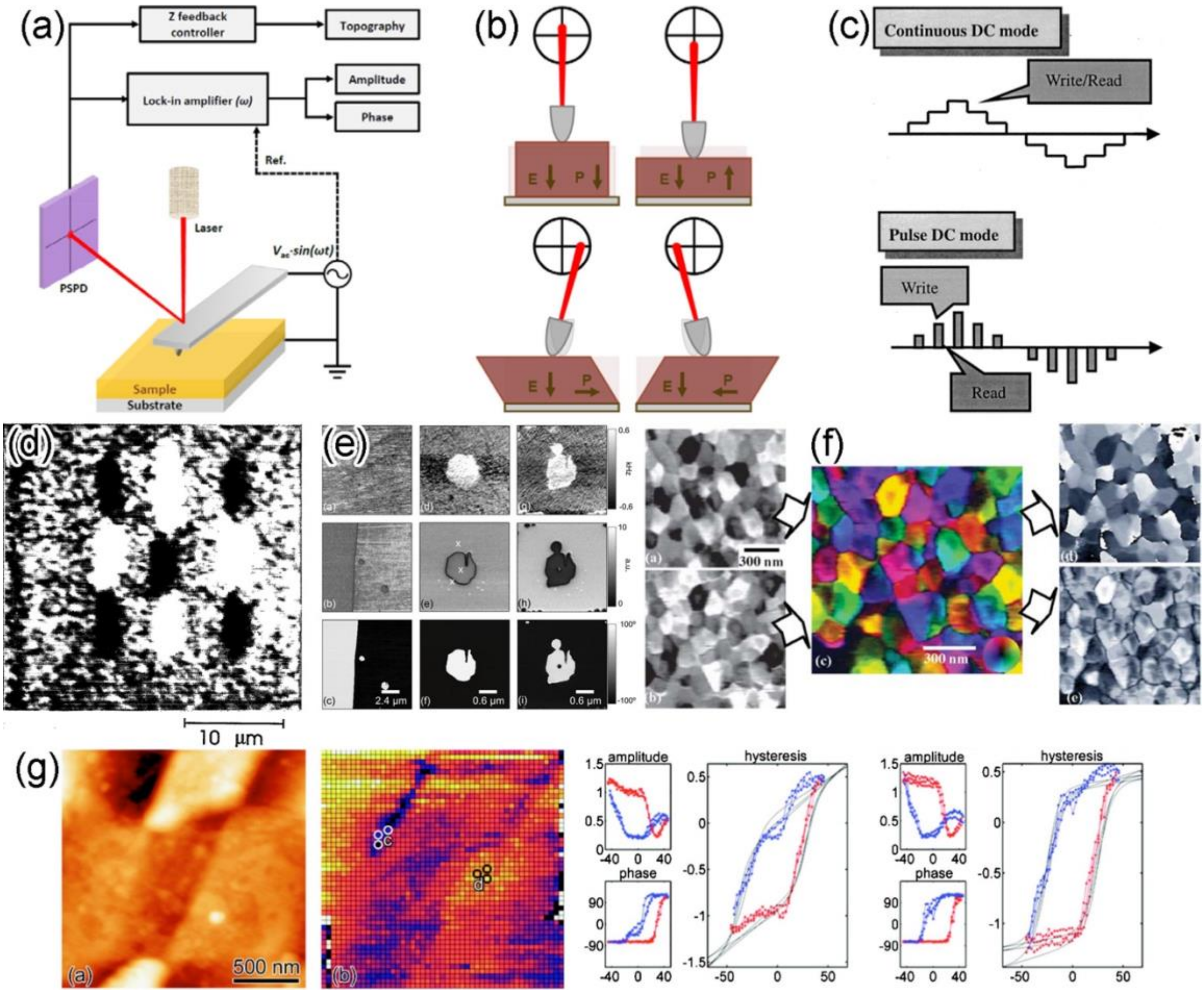

**Figure 4.** Fundamentals of Piezoresponse Force Microscopy: (a, b) schematic representation of PFM response detection in an atomic force microscope. Reprinted from[206] Copyright (2017), with permission from Elsevier. (c) Waveforms used for local hysteresis loop measurements in continuous and pulse DC modes. Reprinted from[190] Copyright (2001), with the permission of AIP Publishing. (d) One of the earliest PFM images of a 1 µm thick PVDF-TrFE copolymer film. Bright and dark spots correspond to regions poled with negative and positive voltages, respectively. Reprinted from[185] Copyright (1992), with the permission of AIP Publishing. (e) Domains in lithium niobate captured in DART mode. Top row: resonance frequency variations, middle row: PFM amplitude, bottom row: PFM phase. Reprinted with permission from[191]. (f) Vector PFM of a $PbTiO_3$ thin film, polarization orientation (angle) is decoded from vertical and lateral PFM signals and visualized as a color-coded map in the central image. Reprinted from[232] Copyright (2006), by permission of Oxford University Press. (g) Topography, PFM scan, and local hysteresis loops in a PZT film. Reprinted from[194] Copyright (2006), with the permission of AIP Publishing.

In recent years, the integration of machine learning algorithms to advance PFM has become a major focus within the scientific community.[5, 244-248] The development of automated approaches is expected to greatly accelerate PFM measurements, improve result interpretation, and align with macroscopic measurement techniques.[23]

### IV.b. Electrochemical Strain Microscopy

Electrochemical phenomena are the basic operational mechanisms of multiple applications including batteries, fuel cells, memories, and other electrochemical systems. While electrochemical phenomena have been typically explored through macroscopic methods such as electrochemical impedance spectroscopy, ionic motion, or electrochemical reactions, they actually occur at the nanoscale and atomic regimes necessitating high-resolution imaging techniques. This can be achieved using a recently developed atomic force microscopic (AFM) technique, known as electrochemical strain microscopy (ESM).[249, 250]

In ESM, when an electric field is applied via the AFM tip, the electric field is concentrated underneath the tip. This triggers ion migration within the material or ionic reactions resulting in a local strain, known as electrochemical strain. When an AC voltage is applied, this mechanical deformation induces oscillations of the tip, which can be detected using a lock-in amplifier that demodulates at the applied AC frequency. By varying the applied voltage and frequency, ESM can provide quantitative information about ionic diffusion coefficients, local ionic conductivity, and electrochemical strain coefficients.[207, 222, 251] The spatial resolution of ESM can typically achieve approximately 20 nm, determined by the size of the AFM tip. However, theoretically, its spatial resolution can be further refined to the order of 3 nm to 10 nm.[252] Indeed, sub-10 nm of the spatial resolution has been reported experimentally.[253] In addition to the typical first harmonic ESM, second harmonic ESM can be used to probe electrochemical dipoles.[254]

ESM has been extensively used to explore local electrochemical phenomena. For batteries, this helps to understand phenomena such as lithium intercalation/de-intercalation, phase transformations, and degradation processes.[207, 249, 251, 255-258] For instance, N. Balke et al. visually showed higher ionic mobility at the grain boundaries in a Si anode[249] and a different ionic mobility of the solid-electrolyte interphase using ESM.[259] Furthermore, ESM images of Li ion concentration and diffusivity with nanoscale resolution were also shown in $LiFePO_4$ and $LiMn_2O_4$.[251, 256] For solid oxide fuel cells (SOFC), as ionic mobility and electrochemical activity at the electrode/electrolyte interface are critical for performance, ESM can provide insights into the ionic transport pathways.[253, 260-262] For instance, in yttrium-stabilized zirconia,

A. Kumar et al. reported oxygen reduction/evolution reactions and higher electrochemical activity at the triple-phase boundary.[253] Furthermore, spatially localized oxygen reduction (ORR) and evolution (OER) reactions were visualized in lanthanum strontium cobaltite (LSCO) (Figure 5c). Their positive/negative bias distributions (PNB and NNB), which can be associated with critical bias ORR/OER activation, exhibit strong correlation with the degree of order of the material (Figure 5d,e).[262] In yttrium-doped barium zirconate, a clear correlation between conductivity and interface structural defects was visually shown by ESM.[261]

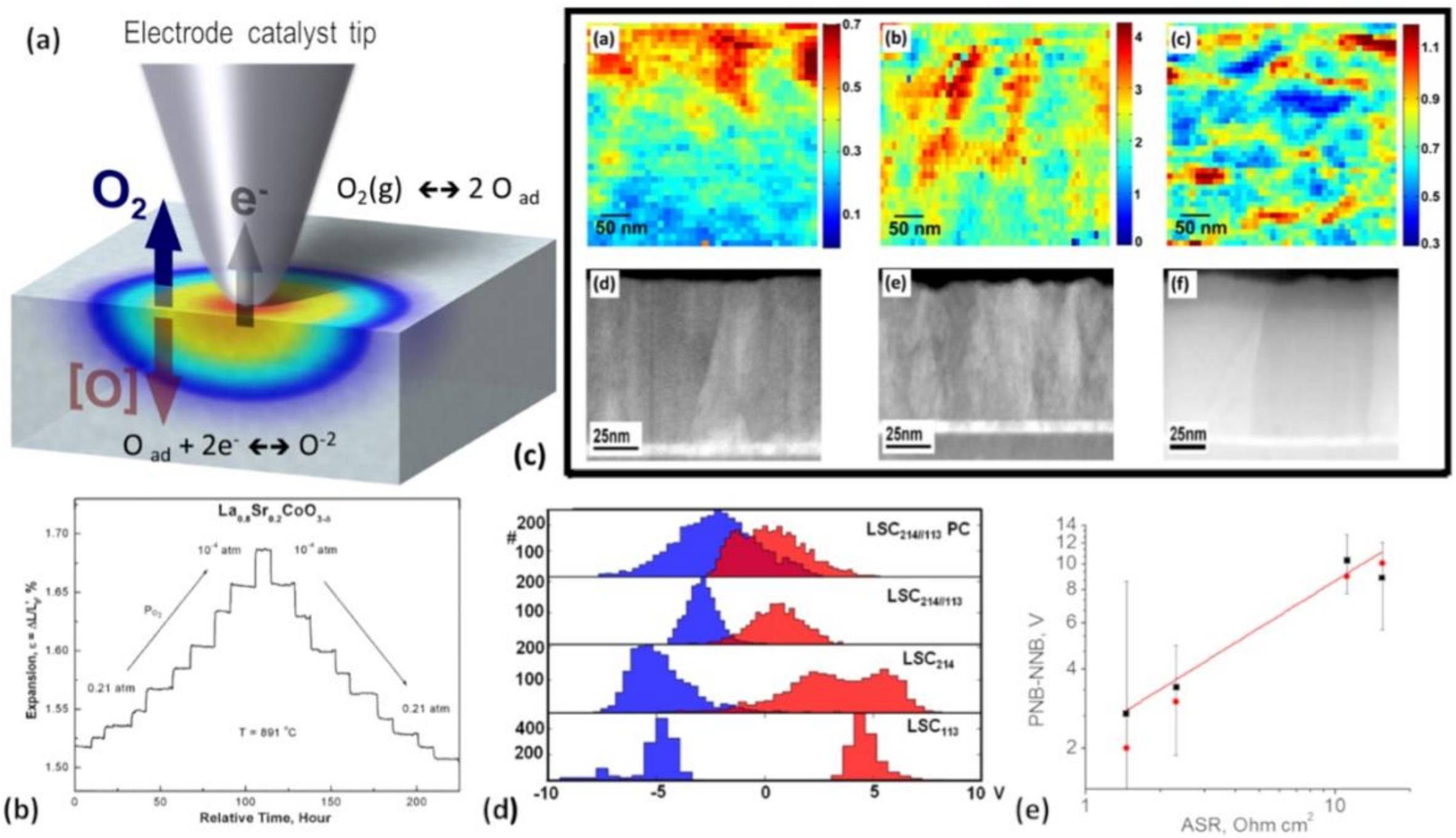

**Figure 5.** (a) Principle of the Electrochemical Strain Microscopy (ESM). Application of the bias to the SPM tip induces the ORR/OER process on the gas-solid interface. (b) Associated change in molar volume due to the Vegard strain results in surface deformation that is detected by an SPM tip with ca. 1 pm limit. (c) ESM map of three LSCO samples with dissimilar surface terminations compared to cross-section electron microscopy images. Note that bright features in ESM images are clearly associated with grain boundaries resolved at the 10 nm level. (d) The measure of electrochemical activity in ESM is the activation voltage, shown here for 4 different electrocatalysts. (e) Comparison of the room temperature ESM signal with the classical area specific resistance measured at 600 °C suggests that the ESM signal can be a proxy for classical electrochemical measurements. Image (a) reprinted with permission from [253] Copyright (2011) Springer Nature Limited. Images (b-e) reprinted with permission from [262] Copyright (2013) American Chemical Society.

In addition to energy applications, there are several reports on ionic information technology materials. For memristive memories, ESM can monitor oxygen vacancy motion or other ionic motions, which are key for the operational mechanism of memristive memories. For instance, P. Sriboriboon et al. reported that memristive switching in $TiO_2$ is associate with irreversible oxygen vacancy motion under a higher voltage regime by combination with ESM and conductive AFM.[263] Oxygen vacancy activity in NiO was reported to be associated with a mechanical stimulus[264] as well as the voltage regime.[265-267] On the other hand, ionic transport phenomena are also crucial for the operational mechanism of organic electrochemical transistors.[268, 269] Therefore, a direct observation of local variations in ion transport[269] was demonstrated in polymer devices by in-situ ESM.

While ESM has proven to be a powerful technique, challenges remain. Translating the measured strain response into quantitative ionic concentrations and diffusivities requires robust theoretical models and calibration standards. The complexity of the material responses and the influence of various factors, such as tip-sample interactions and environmental conditions, need to be carefully considered.

Nevertheless, ESM is a versatile and powerful technique for probing electrochemical reactions at the nanoscale. Its ability to correlate mechanical deformation with electrochemical activity provides unique insights into a wide range of materials and applications. As the technique continues to evolve, it is expected to play an increasingly important role in the development of advanced energy and memory systems, as well as in other fields where electrochemical processes are critical.

### IV.c. Scanning Electrochemical Microscopy (SECM) and EC-cell Research

Within the larger class of SPM, scanning electrochemical microscopy (SECM) is a complex technique used to evaluate local electrochemical behavior at interfaces such as liquid/solid, liquid/gas, and liquid/liquid.[270-277] When this method was initially created in 1989,[278] it completely changed the discipline of electrochemistry by making it possible to conduct in-depth, spatially resolved electrochemical investigations.

The ability to study localized, quick electrochemical responses was made feasible by the discovery of ultramicroelectrodes (UMEs) in the 1980s.[279, 280] UMEs, which are small enough to offer excellent spatial resolution, are essential components of SECM, allowing for precise measurements of electrochemical currents as the tip moves across the material of interest. The basic idea behind SECM is to measure the current at the tip of a UME as it gets closer to and interacts with the sample of interest. Diffusion-limited current, which happens when the rate of electrochemical reactions is regulated by the diffusion of reactive species to the electrode surface, is crucial to the interpretation of SECM signals. In SECM, the feedback mode and the collection-generation mode are the two main operating modes.

When in feedback mode, the sample characteristics influence how the UME tip current behaves. The reduced species at the tip are oxidized at the sample surface when the tip gets close to a conductive sample. This results in a regenerative positive feedback loop that raises the tip current. On the other hand, oxidized species cannot be replenished when the tip gets close to an insulating sample, which results in a negative feedback loop and a drop in tip current. Plotting the variations in current as a function of sample distance from the tip yields useful data on surface conductivity and reactivity; these curves are called approach curves.[278]

There are two variations of the collection-generation mode: sample generation/tip collection (SG/TC) and tip generation/sample collection (TG/SC). When operating in TG/SC mode, the UME tip produces a reactive species that is subsequently gathered by the sample that is held at a particular potential. The sample surface creates a reactive species in the SG/TC mode, which is collected by the tip. These modalities are employed in the investigation of species transfer between the tip and sample, as well as reaction kinetics.[281]

Potentiostats, current amplifiers, piezoelectric positioners, and computers for control and data gathering are common parts of SECM equipment. It is essential to place the UME precisely in relation to the sample, and this is accomplished by combining fine piezoelectric adjustments with coarse motorized positioning. A platinum microwire is typically sealed in a glass capillary and polished to form a sharp tip to prepare SECM probes. High-resolution probes have been developed as a result of improvements in nanofabrication techniques, allowing SECM studies to focus on smaller and quicker events.[282]

The study of solid/liquid interfaces, including electrocatalysis, surface reactivity, and dissolution kinetics, is greatly aided by SECM.[283] For instance, by locally altering the dissolution equilibrium at particular crystal faces, SECM may investigate the dissolution of ionic crystals in aqueous settings and provide quantitative rate data. The analysis of quick response kinetics and differences in solubility between distinct crystal faces is made possible by this great spatial resolution.[284-286]

Surface patterning and microfabrication are further applications for SECM. SECM is used in methods like scanning probe lithography (SPL) to deposit materials or desorb chemical species locally to shape surfaces with micron-sized features.[287-291] This feature is useful for investigating systems attached to small gold clusters or other materials, as well as for assembling nanoscale assemblies.

SECM can be used to test non-conductive surfaces in biological research, including membranes and redox-active enzymes.[292] It is a non-invasive technique to examine ion transport across cell membranes and monitor intracellular charge transfer. Researchers can monitor variations in current associated with intracellular redox processes or ion transport rates by placing the UME close to a cell or membrane. This allows for valuable insights into the dynamics and function of cells. More precisely, they can determine, if the membrane transport rate is uniform or localized and pinpoint the active spots based on the precise transport rate that was recorded.[270]

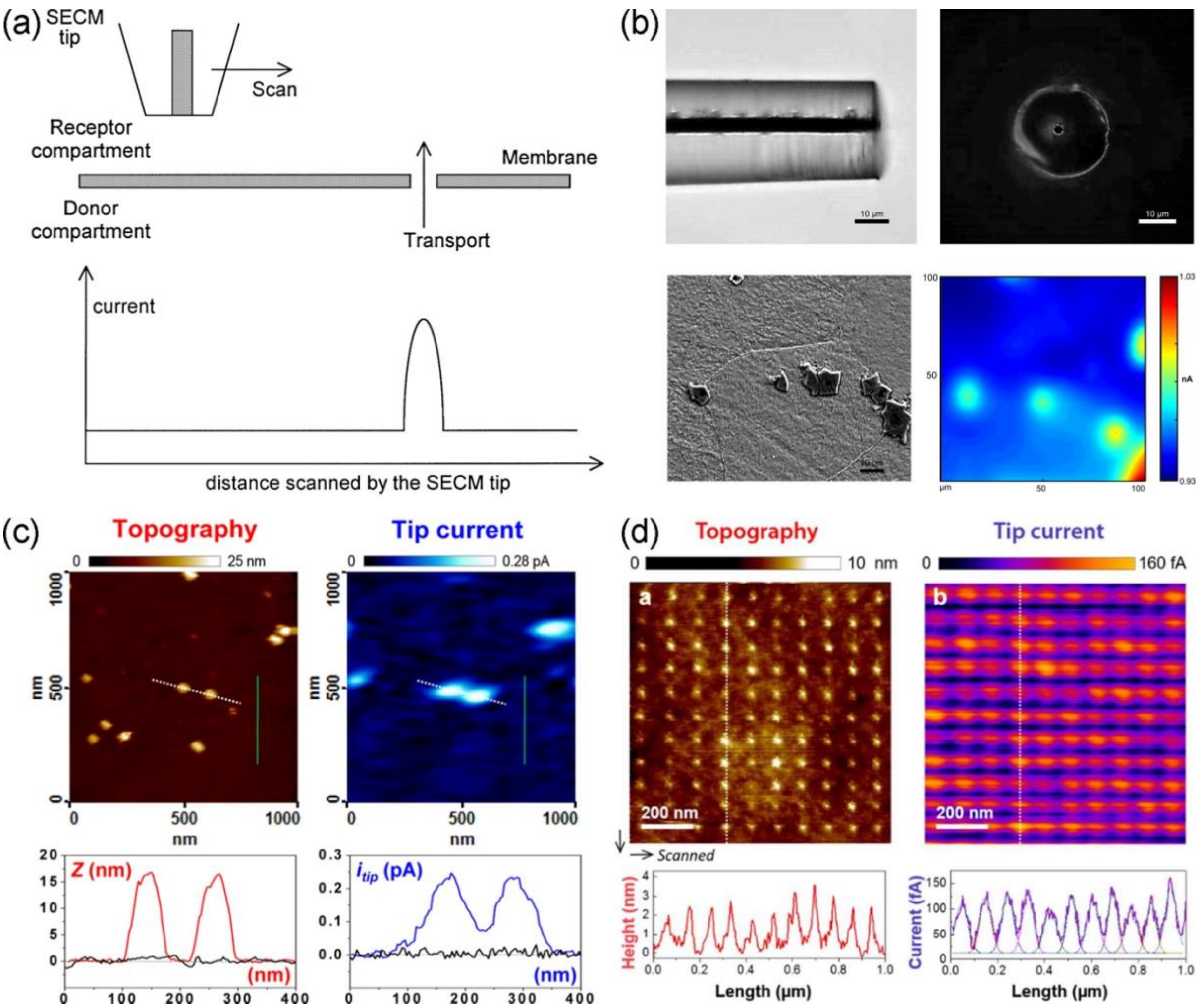

**Figure 6.** (a) Schematic for examining membrane permeability using SECM. The diffusion-limited current map obtained as a function of the tip position can be transformed into a corresponding permeability map of the target interface. Reprinted from[270] Copyright (1999), with permission of Elsevier. (b) SEM images of the carbon microelectrode tip and SECM current image map of Ti-rich inclusions found in high-performance steel proving the role of Ti-rich inclusions in the initiating of localized corrosion. Reprinted from [293] by Gateman S. M.

et al., licensed under CC BY. (c) AFM topography and SECM imaging of a gold surface bearing a random array of approx. 20 nm Fc-PEGylated gold nanoparticles. Reprinted with permission from.[294] Copyright (2013) American Chemical Society. (d) The SECM imaging of the dense Fc-PEGylated gold nanodot array, demonstrating capability for high-resolution SECM measurements. Reprinted with permission from.[295] Copyright (2017) American Chemical Society.

The study of chemical transport kinetics at liquid/liquid and liquid/gas interfaces is another application for SECM. This method can assess chemical transport across monolayer films at air/water interfaces shedding light on the permeability and diffusion characteristics of biological membranes and environmental interfaces. To investigate electron and ion transfer processes, SECM can also examine the electrochemical activity at the interface between two immiscible electrolyte solutions (ITIES).[296]

Even though SECM is widely applicable, there are still obstacles to overcome, like the requirement for extremely accurate probe positioning, the complexity of fabricating probes, and the interpretation of signals in intricate surroundings. To provide thorough insights into electrochemical systems, future developments in SECM will work to improve spatial resolution, refine the fabrication processes of the probes, and combine SECM with other analytical techniques like Kelvin Probe Force Microscopy (KPFM) and ESM.

Using high spatial resolution, SECM is a potent technique for examining the behavior of electrochemistry at different interfaces. In both scientific research and industrial applications, its capacity to furnish comprehensive data on surface reactivity, dissolution kinetics, electrocatalysis, and biological processes renders it indispensable. With the development of technology, SECM's capacity to comprehend and control electrochemical systems will grow, presenting fresh chances for research and development in the fields of biology, chemistry, materials science, and more.

### IV.d. Surface photovoltage measurements by Kelvin Probe Force Microscopy

Surface photovoltage (SPV) is a critical diagnostic tool in both photocatalysis and photovoltaics by providing insights into the surface electronic properties of materials, including charge carrier dynamics, work function, band alignment, surface states, and defect structures.[297] Its application allows for the optimization of material properties and interfaces, ultimately enhancing the efficiency of energy conversion processes in both light-driven chemical reactions and solar energy harvesting devices.

In photocatalysts, which facilitate light-driven chemical reactions at the surface (such as water splitting or $CO_2$ reduction), SPV measurements offer valuable information about charge separation and surface reactivity, since an efficient separation and transfer of photogenerated electrons and holes to distinct surface sites is crucial.[298-300] The charge separation processes are highly intricate occurring over a broad spatiotemporal range from nanometers to micrometers and from femtoseconds to seconds. Consequently, only a small fraction of the charges typically reaches the photocatalyst surface sites.[301] This complexity renders the charge-separation process a rate-determining step in photocatalytic efficiency. A change in SPV signal at the material surface is indicative of charge carrier separation and surface band bending upon light illumination.[302]

In photovoltaic devices, where the goal is to convert light into electrical energy, SPV can play a key role in understanding charge generation, separation, and collection mechanisms. When a semiconductor material absorbs light, it generates electron-hole pairs. The separation of photoinduced charge carriers and transportation to the electrodes directly impacts the photovoltaic device performance. SPV is particularly useful in investigating the behavior of charge carriers at interfaces, such as the semiconductor-electrode or semiconductor-semiconductor junctions in the case of heterojunction cells.[303] The magnitude of the SPV signal provides information about the built-in electric fields at the interfaces, which are essential for driving charge separation. Furthermore, SPV can be used to study band alignment at these interfaces, where misalignment can lead to charge recombination, reducing device efficiency. By monitoring changes in the SPV signal, one can assess the effectiveness of passivation layers or other surface treatments designed to reduce recombination rates and improve device performance.[302, 304]

In both applications, SPV can reveal information about surface defects, interface quality, and band bending, all of which affect the material reactivity and efficiency in light-driven reactions. The SPV signal depends on the type and density of surface states and the presence of adsorbed species or reaction intermediates, which can alter the surface charge distribution. Macroscopic methods, such as X-ray photoelectron spectroscopy, are commonly employed to characterize charge transfer by observing chemical shifts, although these shifts are averaged over the entire catalyst. In contrast, the spatially resolved SPV techniques based on KPFM allow for direct imaging of localized charge separation at the surfaces and interfaces of photovoltaic materials.[305, 306] This method offers nanometer-scale resolution and operates under real-world conditions, providing insights into the real-time behavior of materials during active photocatalytic or photovoltaic processes. Such advancements have significantly improved the design and optimization of various materials, including perovskites, organic photovoltaics, traditional silicon cells, and diverse photocatalytic systems like plasmonic photocatalysts.[307-309]

KPFM, first introduced by Nonnenmacher in 1991,[310] operates by measuring variations in the local contact potential difference (CPD) under ambient conditions. This CPD reflects the difference in Fermi levels between the tip and the sample. The mapping of CPD is performed using a conductive tip positioned slightly above the sample surface. An AC voltage is applied to the tip and the electrostatic response is tracked using a lock-in amplifier. This approach enables two different detection methods: amplitude modulation (AM) mode and frequency modulation (FM) mode. In AM mode, the method tracks variations in the response amplitude. An additional DC voltage ($U_{DC}$) is applied to the tip to achieve accurate measurements. The electrostatic force is minimized when $U_{DC}$ equals the $U_{CPD}$ between the tip and the sample. This condition, $U_{DC} = U_{CPD}$, ensures precise mapping of the surface potential. The FM mode lies in tracking variations in the resonance frequency of the cantilever, caused by electrostatic forces between the tip and the sample.[5] It is important to note, that in KPFM measurements the CPD is influenced primarily by the topmost layer of atoms or molecules, making this technique particularly sensitive to surface chemistry and conditions.

In the *n*-type semiconductors, band bending occurs upwards due to the depletion of majority carriers in the surface region[303] leading to an upward bending of the local vacuum level and an increase in the work function from the bulk to the surface (Figure 7a). When KPFM is utilized, the local vacuum levels of the sample surface and the tip are equalized. Under light illumination (Figure 7b), photoinduced charge carriers are separated by the built-in electric field in the depletion region, driving holes towards the surface and electrons towards the bulk, resulting in a flattened upward surface band bending. The bulk is grounded, maintaining constant energy levels in the bulk under illumination. The flattened upward band bending reduces the surface work function. Since the work function of the tip remains unchanged under illumination, the SPV of the material can be determined by measuring the CPD under illumination and in the dark. For an n-type semiconductor with upward band bending, as illustrated in Figure 7b, illumination increases the CPD and yields a positive SPV. Conversely, for a p-type semiconductor with downward surface band bending,[303] the separation and transfer of photogenerated electrons towards the surface lead to a decreased CPD and a negative SPV.

The measurement of SPV in heterojunction photocatalysts is inherently more complex due to the intricate charge separation processes occurring on multiple surfaces or at interfaces.

Figures 7c and 7d illustrate the schematic measurement of SPV using KPFM on a type II heterojunction photocatalytic system, which comprises an n-type semiconductor surface and a buried p–n junction interface. The measured SPV can be attributed to two main contributions: charge separation across the p–n junction and charge separation within the spatial charge region of the n-type component's surface (Figure 7d). The SPV in the n-type component is positive due to the migration of holes towards the surface. In contrast, the SPV arising from the p–n junction is negative and is expected to be significantly larger than only the n-component.[297] This indicates that the p–n junction primarily governs the charge separation process, resulting in the transfer of electrons to the surface of the n-type component within the photocatalytic system.

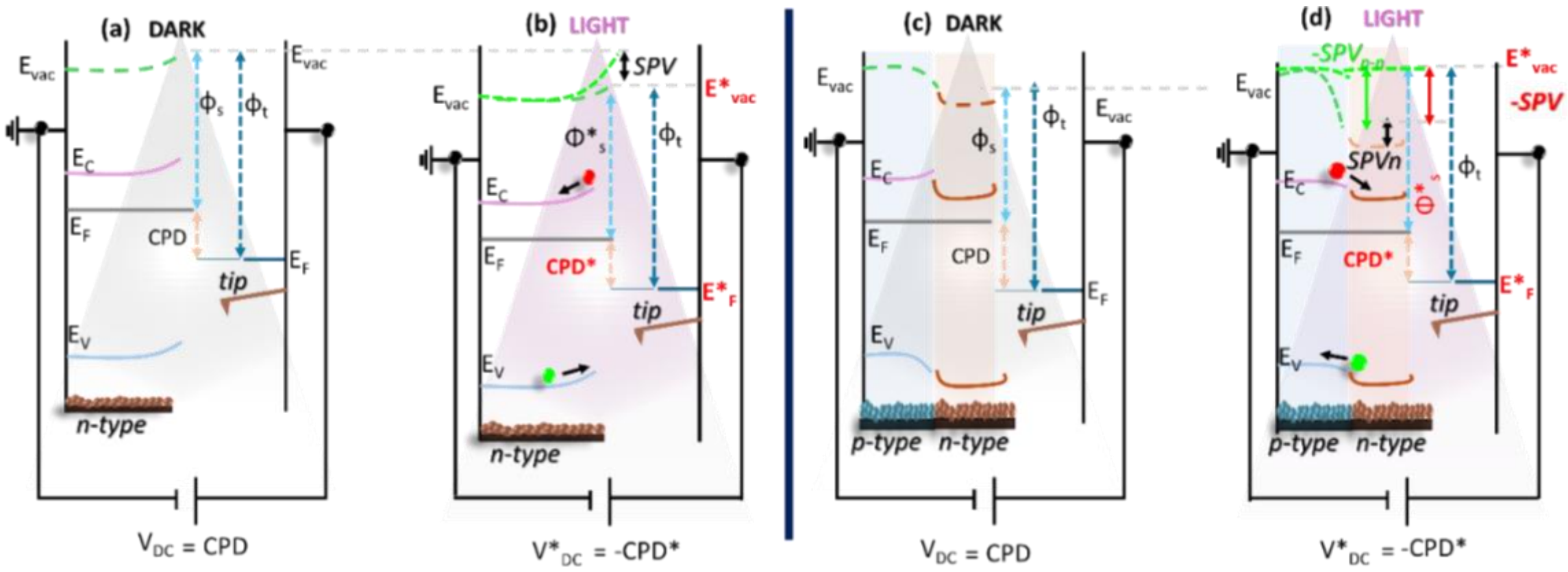

**Figure 7.** Simple schematic of (a) SPV measurements on an n-type semiconductor photocatalyst with upward band bending. (b) under light irradiation. (c) Measurement of CPD and SPV on heterojunction (type II) photocatalysts under dark and (d) under light.

The KPFM measurements under illumination have been effectively used to investigate the spatial distribution of surface photovoltage and to probe nanoscale charge transport at interfaces in solar cell materials and devices in real-time, enhancing our understanding of photovoltaics.[311-313] However, despite its potential, there has been limited application of this technique to the study of photocatalysts. To demonstrate the validity of this method for photocatalysts, Chen et al. conducted SPV imaging and a detailed study on n-type $BiVO_4$ and p-type $Cu_2O$ single crystals revealing that under light illumination, KPFM images brightened for n-type $BiVO_4$ and darkened for p-type $Cu_2O$.[314] It was further confirmed that CPD increases for n-type semiconductors and decreases for p-type semiconductors under illumination.[315] Investigations into the effects of cocatalysts on SPV have demonstrated significant enhancements in charge separation. In particular, the selective photodeposition of $MnO_x$ cocatalysts on {011} facets of $BiVO_4$ resulted in SPV signals increasing by more than three times.[316] Notably, the SPV signal at the {010} facet changed to negative while its magnitude increased substantially. Further experimentation showed that increasing the particle size of nano $MnO_x$ cocatalysts could tune the surface potentials on the {011} facets without affecting the {010} facets. This phenomenon is attributed to electron transfer from $BiVO_4$ to $MnO_x$,[317] which lowers the Fermi level in the $BiVO_4$ bulk, thus altering band bending and enhancing the built-in electric field strength between the two facets. The study of plasmonic photocatalysts, such as those composed of Au nanoparticles (NPs) and $TiO_2$, has further elucidated the complex dynamics of charge separation.[318] Under 550 nm illumination, water oxidation was achieved on Au/$TiO_2$. Dark-state KPFM imaging revealed a belt-like region around the Au NPs, indicating that the surface potential at the Au/$TiO_2$ interface was lower than that of $TiO_2$. This discrepancy is attributed to the formation of a Schottky junction at the interface, which facilitates the accumulation of holes generated by plasmon resonance and promotes the transfer of hot electrons to $TiO_2$ while preventing recombination.[318] By utilizing SPV measurements, it has been observed that nickel (Ni) acts as an electron trap facilitating water reduction, while nickel oxide (NiO) functions as a hole trap in $NiO_x$-loaded $SrTiO_3$ systems.[319] These show the distinct roles of Ni and NiO in enhancing photocatalytic activity by optimizing charge carrier dynamics.[319] KPFM studies on Rh-doped $SrTiO_3$ NPs have demonstrated that the incorporation of ruthenium (Ru) or platinum (Pt) cocatalysts, alongside the use of electron or hole scavengers, significantly improves charge separation efficiency.[320] This improvement is attributed to the modification of the electronic properties of the $SrTiO_3$ surface, which facilitates more effective spatial separation of photogenerated charge carriers. The influence of charge separation at the nanoscale is profoundly impacted by several factors, including the built-in potential of donor-acceptor configurations, the physical segregation of donors and acceptors, and the reversibility of redox reactions. At this scale, where space charge layers are less prominent, these parameters play a critical role in determining the overall efficiency of photocatalytic processes.

In the pursuit of achieving the highest solar cell efficiencies, SPV measurements have become a cornerstone in understanding the mechanisms and charge carrier dynamics of light-absorbing materials in photovoltaics. Nicoara et al. utilized KPFM to study the effect of potassium fluoride post-deposition treatment (KF-PDT) on the surface of Cu(In,Ga)$Se_2$ (CIGS) thin-film solar cells. Through spatially resolved imaging of the surface potential, they observed a marked difference in the electronic properties of grain boundaries (GBs). Specifically, KF-PDT increased the band bending at GBs by approximately 70%, leading to a narrower distribution of work function values at the GBs. This effect was instrumental in the improved efficiency of CIGS solar cells, highlighting the potential of surface treatments in optimizing material performance.[321] Similarly, Lee et al. explored the impact of surface passivation on a wide-bandgap perovskite material (ca. 1.74 eV) using KPFM measurements. By introducing a very thin layer of phenylethylammonium iodide (PEAI) ammonium salt as a passivating agent, they demonstrated significant improvements in the electronic properties of the perovskite surface. The enhanced surface passivation contributed to better carrier dynamics, leading to improved solar cell efficiency. These findings underscore the role of surface engineering in advancing perovskite-based solar technologies.[308] Collectively, these studies exemplify the crucial role of SPV and KPFM in understanding and optimizing the interface properties of materials, driving forward the development of highly efficient and scalable solar technologies.

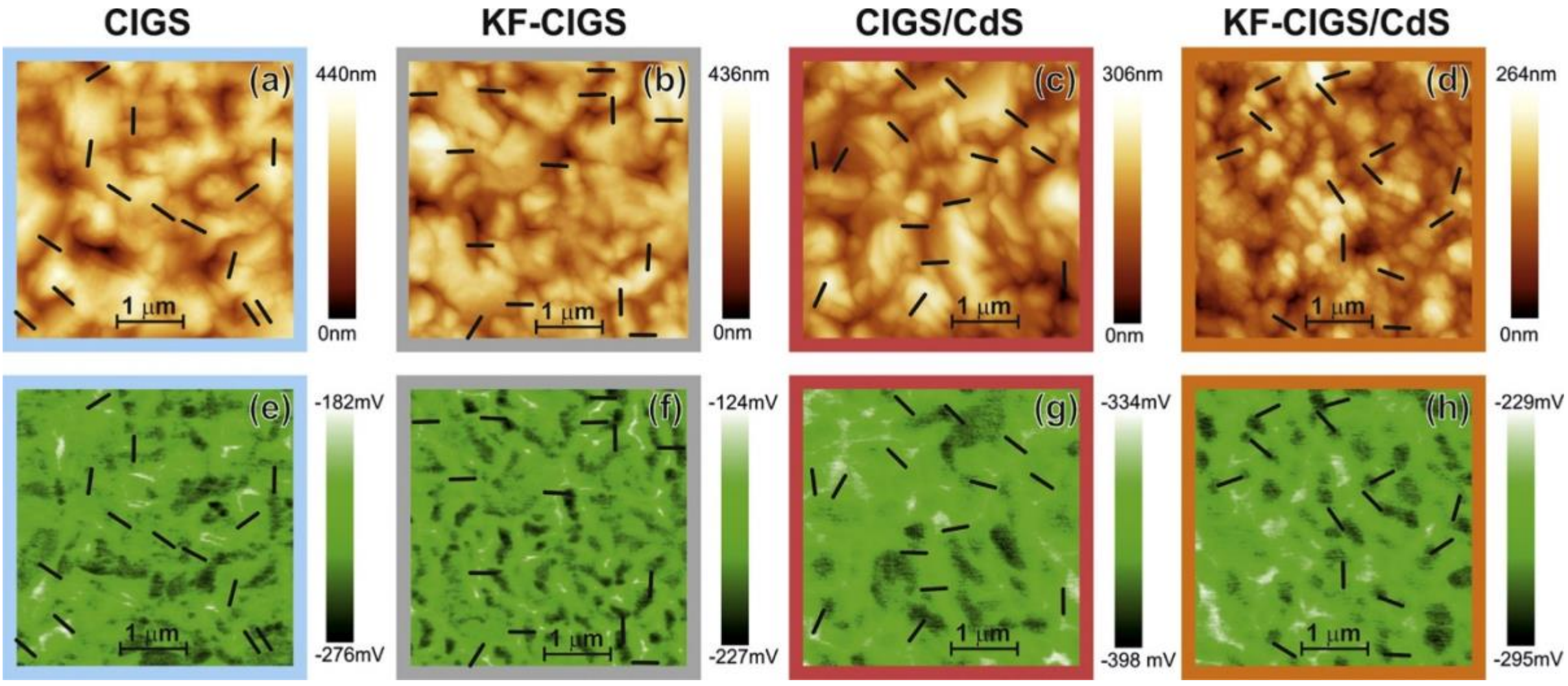

**Figure 8**. KPFM measurements on four samples showing (a–d) topography and (e–h) simultaneously recorded CPD images of CIGS, KF-CIGS, CIGS/CdS and KF-CIGS/CdS. The black lines represent the variations across grain boundaries, based on the topography image for calculation of CPD variation along the line profile. Reprinted from[321] by Nicoara N. et al., licensed under CC BY.

SPV plays a crucial role in enhancing the efficiency of solar cells. By homogenizing the SPV of the light-absorbing materials and modulating its magnitude, we can significantly reduce interfacial barriers and improve charge injection efficiency in devices. This opens up exciting opportunities to explore key questions, such as how SPV uniformity minimizes interfacial recombination and boosts carrier injection and the role of external electric fields or polarization in optimizing junctions and interfaces. Furthermore, investigating the Schottky barrier height via cross-sectional SPV distributions, using techniques like KPFM, can provide a deeper understanding of these barriers. Chemical methods such as doping and the addition of interlayers will allow precise tuning of SPV distributions, enhancing interfacial charge transfer. By controlling these factors, we can manipulate polarization effects induced by external fields, ultimately improving device performance and enabling the scalable, low-cost manufacturing of next-generation solar technologies.

### IV.e. Conductive Atomic Force Microscopy

Conductive AFM (CAFM), which uses a conductive tip to scan the surface of a material while simultaneously measuring the local current flow by applying bias voltages, combines the high-resolution imaging capabilities of standard contact AFM with the ability to measure electrical properties through the bulk at the nanoscale.[322, 323]

In the imaging mode, CAFM uses a conductive tip scanning the sample surface to record topography; in the meantime, a constant voltage is applied between the tip and the sample substrate (or bottom electrode) to measure local current flow through the tip-surface junction, allowing for the mapping of electrical conductivity at the nanoscale. The topography offers the usual insights into physical structures such as surface roughness, grain boundaries, etc., but can also induce a capacitive current signal, which changes its sign depending on the scan direction, in which case reducing the surface roughness by e.g. polishing might be required.[324] The conductivity map shows the local electrical properties highlighting areas of high and low conductivity. Correlating the conductivity map with topography allows us to gain crucial insights into the properties and performance of the materials under study. For example, in semiconductor materials, the conductivity variation may relate to different materials phase, doping levels, or defect sites. This method has been widely used to study local electrical properties of photovoltaic perovskites,[325] organic heterojunction, 2D materials, metal oxide, carbon nanotube,[326] dielectrics materials,[327] etc., revealing grain-to-grain conductivity variations,[325, 328] grain boundary effects on conductivity,[329] or breakdown and leakage.

In addition to mapping, CAFM can also operate in spectroscopy mode, by measuring current-voltage (I-V) curves at each spatial location. In IV spectroscopy mode, a low-frequency sweeping voltage bias is applied to the conductive tip in contact with the sample surface, while recording the flowing current. Upon finishing the sweep at one location, the tip is moved to the next one on the grid, thus building a library of local conductive behaviors. IV-spectroscopy measurements can also characterize temporal behavior of samples, when the current response is recorded as a function of time under a constant or varying voltage bias, or other factors such as light illumination or varying temperature. These temporal current spectroscopy measurements can offer information[330-333] on the dynamic processes such as charge trapping/detrapping, ionic migration,[334-336] and relaxation. These are important characteristics for electronic and optoelectronic materials, such as hybrid perovskites.[337] The shape and characteristics of IV-curves carry information on the local conduction mechanism, barrier height, doping level, and interface properties.[338] A typical FORC-IV unipolar voltage waveform, consisting of sequential triangular pulses with gradually increasing peak bias, is shown in Figure 9b, inset, together with the current response and IV curves, collected in some locations of a $BiFeO_3$-$Co_2FeO_4$ (BFO-CFO) sample. In this case, the IV response is non-hysteretic, although non-linear. The appearance of hysteresis in IV-curves can indicate local Joule heating, RC-response, ion migration, Faradaic processes, charge trapping, or material degradation during voltage sweeping.[339-341] The first two reasons are extrinsic to the studied material and have to be eliminated: Joule heating by limiting the current to below ca. 100 nA (or dissipated power to below 100 nW), and second, by adjusting the voltage sweep rate such that a steady-state current is measured after the RC-decay has leveled off (more on this below). Two examples involving controlled ion migration and Faradaic processes, NiO and $TiO_2$, are shown in Figures 9c,d. In the case of NiO,[342] resistive switching hysteretic behavior is induced by changing the ambient from argon or dry synthetic air to humid air (with water electrochemical splitting providing hydrogen ions that enhance conductivity). With $TiO_2$,[343] a Ti-coated probe provides a source of Ti ions making the measured IV-curve stable, unlike the case of a Au-coated probe (Figure 9d).

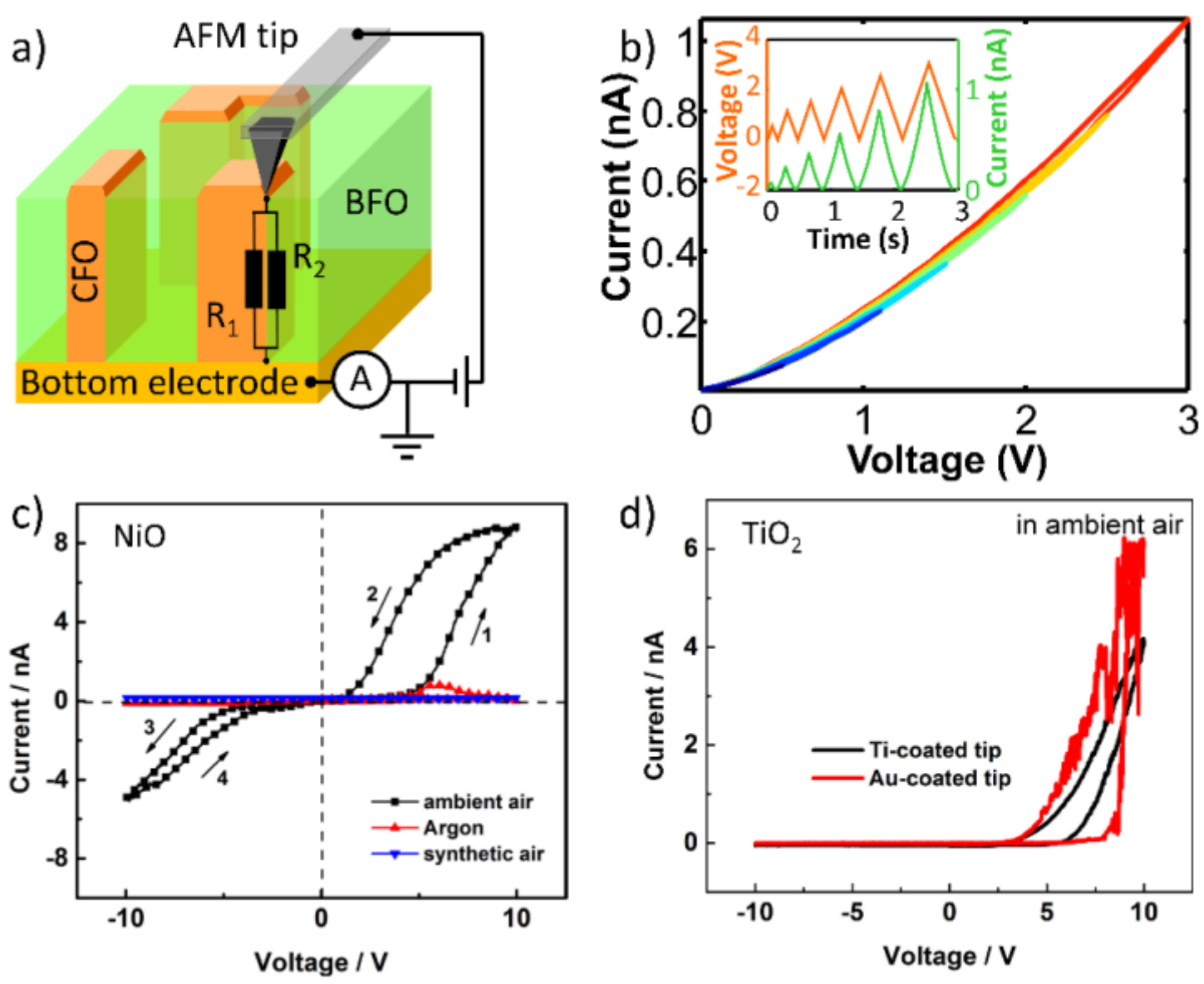

**Figure 9.** Hysteretic IV-spectroscopy. a) A schematic of FORC-IV measurements on a $BiFeO_3$ matrix with imbedded $Co_2FeO_4$ islands showing the interfacial current flowing through both materials, modeled as resistors. b) The inset shows a FORC-IV response, collected on some locations of the BFO-CFO sample in response to the voltage waveform shown in the inset. The response was non-hysteretic, and non-linear. a) and b) Adapted from ref.[344] by Belianinov A. et al., licensed under CC BY. c) Dissimilar IV spectra collected from a NiO sample in various ambients demonstrating the importance of Faradaic processes and control over experimental parameters. Reprinted with permission from ref.[342] Copyright (2018) American Chemical Society. d) IV spectra collected on a $TiO_2$ sample with different metal coatings. Reprinted with permission from ref.[343] Copyright (2018) Elsevier.

Correlating IV-spectroscopy with image characterization furthermore allows us to gain a deeper understanding on how local structures influence electrical behavior. In dielectric materials, IV-measurements can determine the breakdown or leakage voltage at specific locations. Linking this with morphology and roughness from topography images can help to understand the breakdown mechanism and improve materials. The multidimensionality of IV-spectroscopy data calls for the use of statistical and clustering algorithms to visualize, denoise, attribute the data, and fit them to appropriate physical models.[345, 346] This process is exemplified by the BFO-CFO FORC-IV data analysis shown in Figure 10. While both the BFO matrix and CFO islands are insulating at low voltages, their interface is conductive[347] (Figure 10a-b). Using Bayesian Linear Unmixing[348] on the 4-dimensional FORC dataset allowed for the splitting of the data into four loading intensity maps and corresponding endmember IV-curves (Figure 10c-d). Thus, the plethora of local conductive behaviors was expressed as only four fundamental conductive curves: Ohmic conductance of most CFO islands; Schottky barrier transport of the CFO island cores; low non-linear BFO matrix conductivity; and hysteretic, memristive behavior of the interfaces. The endmembers can be fitted to specific models, yielding the transport-governing parameters.[344]

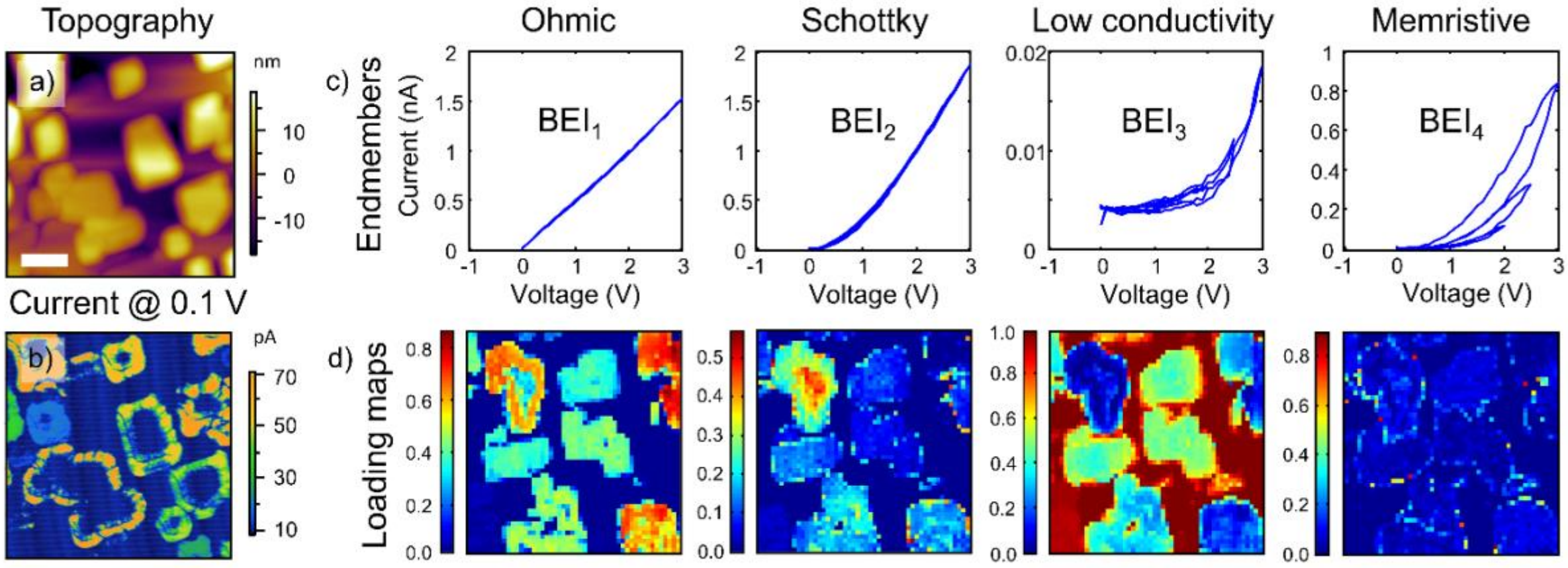

**Figure 10.** The use of statistical tools to visualize and deconvolute the multidimensional IV-spectroscopy datasets. a) BFO-CFO topography with CFO islands protruding above the BFO

surface. Scale bar is 100 nm; b) corresponding current map at low voltage, showing conductivity at the interface; c) Bayesian Linear Unmixing endmembers and d) corresponding loading maps of a FORC-IV dataset showing four distinct types of local conductive behavior. Panels a) and b) were adapted with permission from ref.[347]; Copyright (2013) American Chemical Society. Panels c) and d) were adapted with permission from ref.[345]; Copyright (2013) American Chemical Society.

In photovoltaic materials, correlating IV-measurements with image data can help identify sites where recombination losses occur, and ion migration is higher. This can provide information for targeted passivation to improve the photovoltaic efficiency. Combining spatial image data with IV-spectroscopy data requires sophisticated data analysis approach to correlate them and extract physics, where ML algorithms play a critical role. ML can be also used to guide the experimental design for faster, more reliable data acquisition, which is critical when characterizing complex systems. Traditional IV-spectroscopy measurements can be slow, since the large sample resistance and any parasitic cantilever-surface and in-line capacitance add a long RC constant "tail" to the current response (Figure 11a). As an alternative, Somnath et al.[349] introduced general mode IV (G-IV), which uses a 200 Hz AC voltage excitation (Figure 11c), with a ms pulse duration and allowed them to collect more than 120,000 IV curves on a 256×256 grid within only 17 minutes. The recorded IV-curves have a large oval-shaped hysteresis (Figure 11d,e), which can be removed by using Bayesian inference reconstruction. The resultant curves can be used to extract local capacitance, polarization, and dielectric constant values.

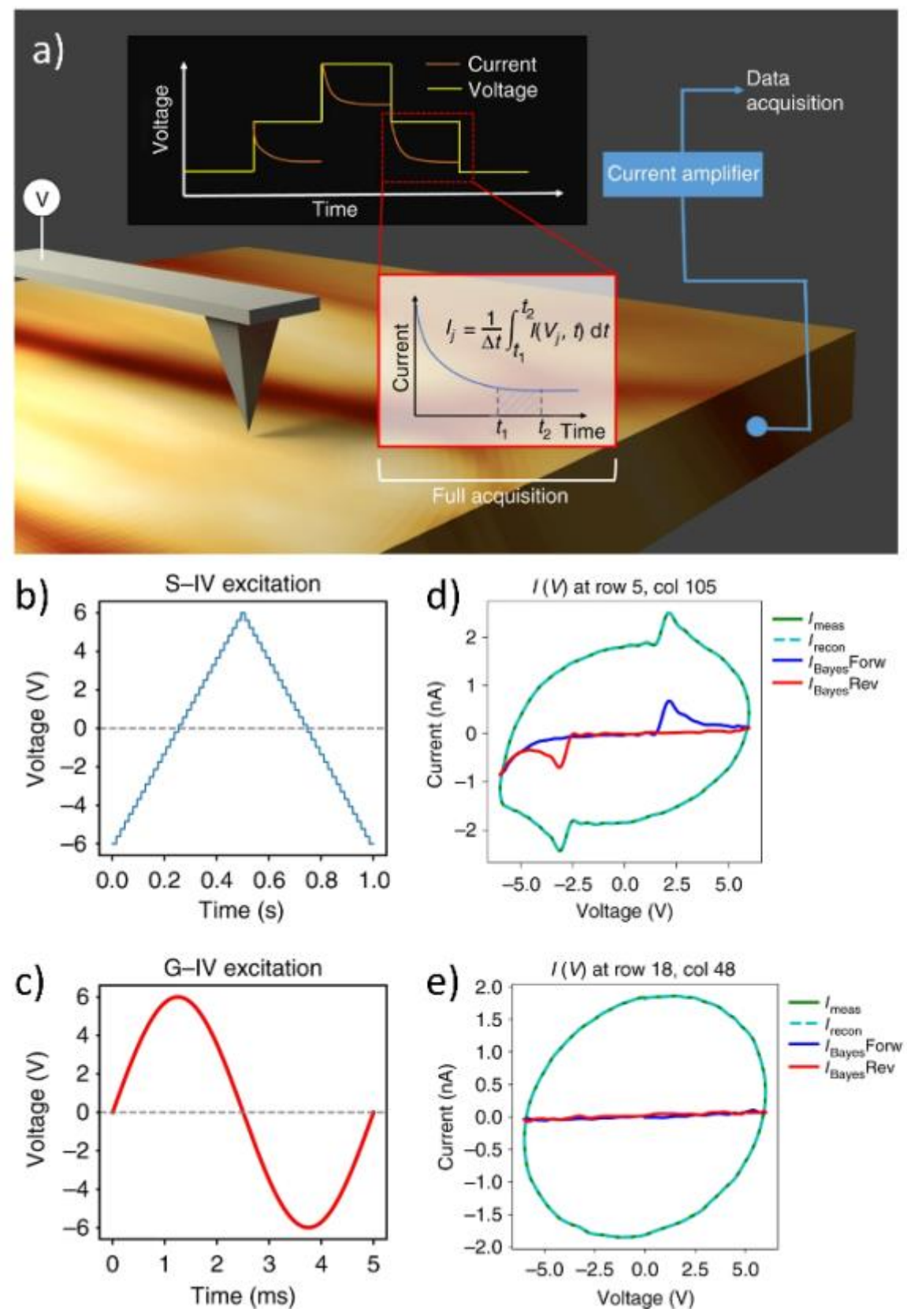

**Figure 11.** Ultrafast IV spectroscopy by Bayesian inversion. a) Measurement schematic with a stepped voltage waveform and current response to it. The response suffers from a long RC-discharge-related settlement time. b) A traditional slowly varying stepped voltage waveform compared to c) a general mode AC excitation waveform. d) A G-IV response measured on a nanocapacitor and e) on a ferroelectric film. The large oval hysteresis of the measured curves is due to the RC contribution, which is filtered out by Bayesian inference algorithm, yielding a flat dielectric response of the film, vs. spiked response of switching nanocapacitor. Reprinted from ref.[349] by Somnath S. et al., licensed under CC BY.

Recently, Liu et al.[350] demonstrated that ML can be used to speed up data acquisition in IV-spectroscopy by automated selecting regions of interest. Preliminary cathodoluminescence imaging of a photovoltaic film revealed higher activity in the grain boundaries, as compared to the grains. Therefore, it was important to explore the electric behavior of the grain boundaries, rather than the rest of the sample. An ML algorithm was used to detect grain boundaries from an AFM topographic image and then automatically perform IV spectroscopy at the respective locations (Figure 12b).

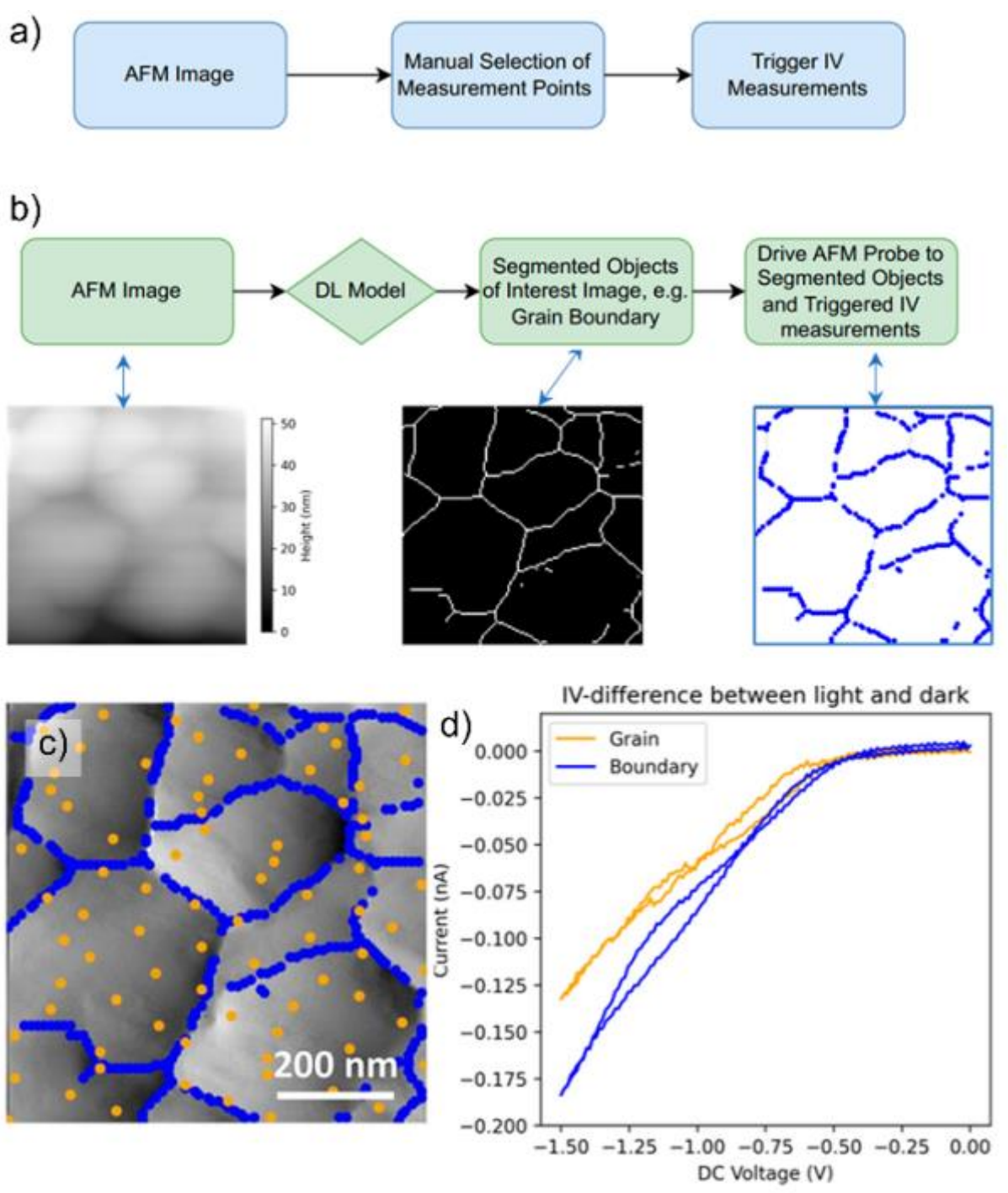

**Figure 12**. Automating IV-spectroscopy. a) A traditional IV-measurement algorithm, when the experimenter manually selects the locations at which to perform measurements. b) Automated SPM uses an acquired topography image and a ML algorithm to detect regions of intertest (in this case – grain boundaries) and then automatically navigate the AFM tip to these locations to perform measurements. c) A topographic map of a photovoltaic material with automatically detected grain boundary locations (blue dots) and grains (orange dots). d) IV-curve difference between illuminated and dark sample measurements at the grain boundaries and grains. Reprinted with permission from ref.[350] Copyright (2023) American Chemical Society.

The use of ML and AI is expected to accelerate data acquisition, analysis and interpretation much more significantly then simply identifying grain boundaries and performing measurements at them. By simple prompt requests,[351] GPT4 can generate Python codes to perform specific measurements (e.g. move the AFM tip along a flower-like shape, run band excitation measurements with required parameters), rapidly visualize any part of the data, select extreme points, etc. Coupled with rapid collection of conductive behavior libraries, ML and AI methods can significantly advance materials characterization and discovery.[352]

### IV.f. Needs and challenges

The development of SPM techniques and instrumentation over the last three decades has been driven by several key considerations: reducing noise, increasing quantitativeness for classical SPM methods, broadening the range of environments for measurements, and enhancing ease of use. Efforts to improve the signal-to-noise ratio enabled clearer and more accurate imaging at the nanoscale. Increasing quantitativeness has involved refining measurement techniques to obtain more precise and reliable data, crucial for linking observed phenomena to material properties. Additionally, advancements have been made to perform SPM measurements in a variety of environments, including variable temperatures and controlled atmospheres, allowing researchers to study materials under conditions that closely mimic real-world applications. Comparatively, the adoption of multiple new SPM modalities proposed over the last 20 years remains fairly limited. This is primarily due to the complex image formation mechanisms associated with these modalities, which complicate the interpretation of data in terms of material functionalities. The complexity in data interpretation arises from the intricate interplay between the tip and the sample, making it challenging to directly correlate SPM images with specific material properties. As a result, while innovative SPM techniques offer potential insights into material behavior, their practical utility is often constrained by the difficulty in extracting meaningful and quantitative information.

Using SPM for rapid materials characterization as part of self-driving labs requires continuous progress in these applications but will also necessitate new developments. Integrability with other automated tools is crucial, ensuring seamless operation within a fully automated experimental setup. Enhancing the quantitativeness of SPM techniques will be essential for providing accurate, reproducible data that can be reliably used for material assessment. Moreover, the integration of advanced ML methods will be paramount for both control and data analytics. ML can automate the interpretation of complex SPM data, improve the precision of measurements, and enable real-time decision-making in the synthesis and characterization processes. By developing robust ML frameworks, SPM can be leveraged to its full potential accelerating the discovery and optimization of new materials. In the following sections, we analyze these requirements in detail exploring the necessary advancements to integrate SPM into the next generation of automated, self-driving labs.

### V.a. Techniques

The distinctive characteristic of the SPM techniques is their capability to probe materials structure and functionality on small length scales, making them ideal tools for the characterization of the combinatorial libraries and small volumes of materials created by the microfluidic, pipetting, or dip-pen lithography-based methods. However, with very few exceptions, the fabrication methods are expected to provide systems with non-uniform surface topography that can couple into the measured signals. These effects are especially significant for the contact mode methods. For SPM-based mechanical measurements and conductive measurements, signal scales with the contact area.[353, 354] For PFM the fundamental signal is independent of topography;[355] however, secondary couplings due to the changes in the resonance frequency of the cantilever are possible.[191] For the non-contact methods, topography can couple to the measured signal via the gradients of the capacitive forces.[356-358] However, these effects are considerably weaker for the detection signals based on differential signals such as photovoltage measurements.

Correspondingly, the key task for the SPM-based materials discovery is the quantification and deconvolution of topographic effects. This can be accomplished via instrument-based strategies such as probe calibration, correction methodologies, or sample design. The calibration of the probe is the first step on the way to decoupling the target response from the topography impact. The complexity of this task arises from multiple factors. Firstly, the observed topographical signal is a result of a convolution between the tip with unknown geometry and the unknown topographical features of the sample surface ("blind deconvolution"). Changes in the tip geometry – due to uncontrolled degradation, contamination, or other factors – introduce alterations in the observable signal. In the case of "blind deconvolution," isolating true surface features from artifacts introduced by the evolving condition of the tip during scanning is particularly challenging. Another complication arises from the mechanical nature of the SPM detection system. The interaction of the probe with the surface is inherently nonlinear due to the finite hardness of the surface, which often varies within a single scan, adhesion forces, surface adsorbates, and contamination, water meniscus in the tip-surface contact, etc. These variations are also reflected in the measured signal. From a technical point of view, the radical solution for the probe calibration can be achieved by integrating scanning electron microscopy (SEM) and SPM microscopy. In a 'fusion scope' a single surface region can be sequentially examined by both methods yielding complementary data that enhance calibration accuracy. SEM can also be used to monitor the condition of the probe, allowing for direct observation of its changes throughout the experimental process.

Another approach for achieving artifact-free SPM measurements lies in developing advanced methods that account for multiple influencing factors observed in various SPM experiments. While capturing the target functionality from a single measurement is challenging, complementary data from multiple SPM experiments can help isolate the target signal from topographical crosstalk. For instance, degradation effects such as tip doubling appear in the measured signal as scanning direction-dependent artifacts, which can be detected by analyzing topography scans from different probe passes (e.g., left-to-right, right-to-left, bottom-to-top, top-to-bottom) to reveal underlying signal distortions.[359] Variations in surface stiffness and adhesion forces, reflected in the phase of SPM tapping mode, can also be observed in force-distance curves. Analysis of the ensemble of diverse SPM signals enables more effective separation of the true signal from probe-related and topographical artifacts.[360] AI-based approaches hold particular promise for handling this intricate analysis, enabling the integration of multiple proxy signals for unsupervised and semi-supervised interpretation.

The third way to enhance the explainability and quantifiability of the measured response is to prepare sample systems with predefined surface characteristics. An effective strategy for mitigating topographical artifacts in SPM measurements is the preparation of sample systems with controlled and predefined surface characteristics. One promising method involves creating a droplet library using automated pipetting robots.[97, 361] In this approach, precursor solutions are dropcast onto a glass substrate in an array format. The use of automated robotics ensures precise control over droplet volume, spacing, and composition, enabling reproducible and systematic studies, as demonstrated in studies like those by Agthe et al. (2013)[362], which highlight the role of drop-casting in creating ordered nanostructures, and Thete et al. (2009)[363], where nanoprofiling was integrated with SPM to examine microarray topographies. Following dropcasting, the samples are subjected to controlled annealing to remove solvent residues and facilitate the formation of a stable final product. For instance, substrates such as mica can be employed to achieve atomically flat surfaces after annealing providing a minimal topographical baseline for subsequent SPM characterization, as shown in the development of superhydrophobic films.[364] In specific applications, resist coverage followed by a peeling process can be used to generate inverted droplet libraries, where the flat surfaces result from the underside of the droplet interface. Such engineered surfaces not only minimize the coupling of topographical features into the measured signals but also provide a well-defined platform for studying material properties across combinatorial libraries, as seen in studies involving orthogonal polymer formations and nanoparticle activity assessments.[365, 366]

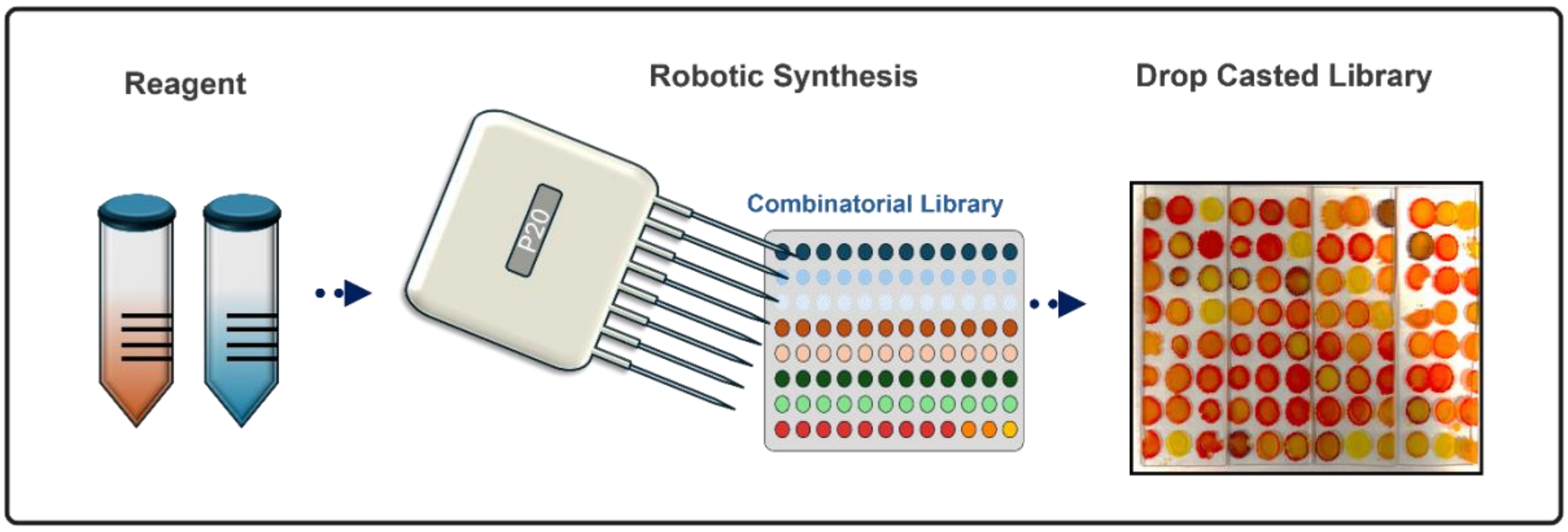

**Figure 13**. High throughput combinatorial droplet library schematic.

### V.b. Platforms

Enhancing the capabilities of SPM microscopes for faster, less noisy scanning has been an ongoing, iterative process since their invention in 1986. Despite significant progress, there remains ample opportunity to further enhance microscope capabilities. Here, the first challenge lies in accelerating the scanning process itself. In most cases, the scan rate (number of lines per second) does not exceed 1 Hz. As a result, the time required for single scan acquisition can take anywhere from a few minutes to over an hour, depending on the scan size and the required spatial resolution. While this mesoscale time range is sufficient for characterizing static properties, in many cases, the duration of a single scan is insufficient for investigating time-dependent processes. The upper limit of the scanning rate is often dictated by the sample surface roughness. In SPM the feedback loop often relies solely on a PID controller to maintain tip-sample interaction, but the reactive nature of this method constrains a scanning speed. One potential solution is to incorporate real-time topography forecasting into the feedback loop, enabling anticipation of surface changes. However, developing and implementing models capable of accurate real-time topography prediction present significant challenges.

The second trend is the expansion of measurement condition ranges. Modern material investigation and characterization often require measurements under diverse conditions, such as varying atmosphere, temperature, humidity, illumination, magnetic fields, and liquid environments. While many instruments already offer these capabilities, further expanding the range of accessible conditions, along with providing user-friendly programming interfaces for their remote control, is a crucial step towards enhancing the integrability of SPM into broader characterization workflows, particularly within automated laboratory complexes. Precise control of experimental conditions is vital for the reproducibility of SPM investigations.

A key global challenge lies in the integration of SPM machines into automated laboratories for high-throughput synthesis workflows. As outlined in Section II, further accelerating material design demands a shift towards smaller length scales in materials production to enhance throughput, alongside the incorporation of microscopical characterization techniques that offer active feedback for material optimization. SPM, capable of providing nanoscale insights into a diverse array of material properties, is emerging as a critical tool for closing the synthesis-characterization loop.

As a global challenge, we emphasize the integration of the SPM machine into automated robotic laboratories for high-throughput synthesis workflows. As it was described in Section II, further acceleration of material design requires a shift towards smaller length scales in materials production to increase throughput, along with the integration of sample characterization techniques to provide active feedback for material optimization. The relatively low cost of SPM machines, their adaptability to diverse operating conditions, and minimal sample preparation requirements make their integration technically straightforward. Furthermore, many SPM microscopes already offer a high degree of autonomy including capabilities for automated tuning. Recent advancements in AI-based optimization of scanning parameters have even enabled autonomous adjustments for advanced measurements, further enhancing their suitability for high-throughput and automated workflows.[367, 368]

In summary, we anticipate significant technical advancements in SPM techniques over the coming years including increased autonomy, expanded remote control capabilities, and a broader range of operational conditions. These developments will enable the design of SPM systems seamlessly integrated into autonomous, high-throughput synthesis labs, positioning SPM as a pivotal tool for rapid and efficient material characterization.

### V.c. ML accelerated workflows

At present, the mainstream of the SPM field are single-probe systems, which make detection inherently sequential—capturing data point by point. This sequential nature poses significant challenges for exploring large samples, such as combinatorial libraries or those with highly diverse microstructures. This limitation is especially pronounced in spectroscopic measurements, where each point requires extended acquisition time to capture detailed information.

In the case of combinatorial libraries, a straightforward approach for automating exploration is grid search, where the library is scanned point by point with a constant step size across nearly every composition. Recently, a grid-based automated experiment was employed to investigate the evolution of ferroelectric properties within the $Sm_xBi_{1-x}FeO_3$ combinatorial library (Figure 14a-d).[369] While this approach provides the most comprehensive information about property evolution across the compositional range, a constant-step grid search can be inefficient and redundant. The evolution of the target property within combinatorial spaces is often non-uniform, making it advantageous to concentrate measurements in critical compositional regions, such as near phase transitions, where higher-density data is valuable. In other regions, fewer observations may be enough to get a comprehensive picture. The decrease in the number of measurements and automatic prioritization of the most important regions is a natural way to enhance the efficiency of the automated exploration and characterization.

The Gaussian process Bayesian optimization (GP-BO) methods have become a golden standard for the automated optimization and exploration across fields ranging from materials synthesis to automated X-ray scattering and probe microscopy.[370-373] The idea of Gaussian processes (*GPs*) lies in the reconstruction of some scalar property dependence $f(x)$ over a low-dimensional parameter space $x \in R^N$ by the number of noisy observations $y(x_i),\ i = 1 \dots n$.[374] The low-dimensional nature of compositional spaces makes GP particularly effective models for automated exploration within combinatorial libraries. In result of such reconstruction, an algorithm provides both the predicted function value (mean function) and its uncertainty within the explored range. Using the mean function and uncertainty estimates, Bayesian optimization constructs an acquisition function that identifies the next location (composition or object) to be explored maximizing progress towards the predefined experimental objective. In other words, GP-BO autonomously defines the optimal experimental trajectory to efficiently achieve the set objective.

The standard GP-BO method is a purely data-driven approach that does not incorporate prior physical knowledge limiting its efficiency for investigating complex physical systems such as combinatorial libraries with convoluted compositional dependencies of target properties. However, there are several opportunities to improve the efficiency of automated experiments by integrating additional information. These include developing physics-informed GP kernels and mean functions (as opposed to the default zero-function mean in standard GP) and embedding physical knowledge through boundary conditions. Incorporating physical information in the form of a tailored mean function (*structured GP, sGP*) has proven particularly effective in recent years.[244, 375-377] In this approach, the parameters of the mean function are defined probabilistically allowing the algorithm to dynamically adjust them during the exploration process for improved performance.[244] The sGP approach is particularly effective for reconstructing property dependencies with sharp discontinuities, such as those occurring at phase transitions. These irregular drops are challenging to capture accurately using standard GP methods. [378] The sGP approach has demonstrated its efficiency for the hypothesis learning problem, where theoretical models or mechanisms are identified from a set of candidates based on a limited number of observations.[244, 375]

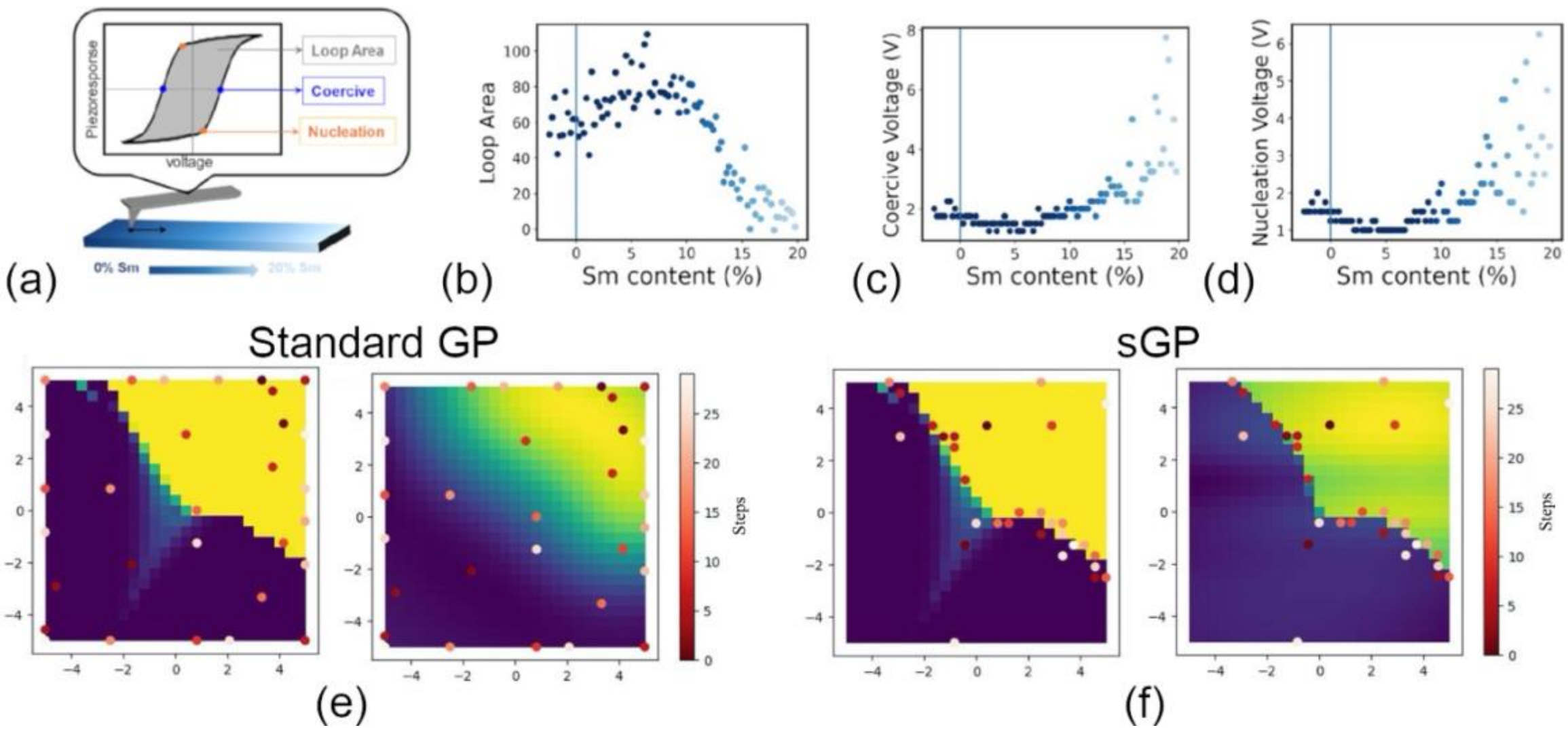

**Figure 14.** (a) Scheme of descriptors extracted from local PFM hysteresis loops and (b-d) their evolution across an Sm-BFO combinatorial library. Reprinted with permission from.[369] Copyright (2024) American Chemical Society. (e,f) Active learning of magnetization behavior in a 2D Ising model. Ground truth (on the left) vs. predictions (on the right) from the points sampled using the (e) standard GP and (f) sGP. Reprinted from.[378] by Ziatdinov M. et al., licensed under CC BY 4.0.

Another approach to enhance the optimization of properties with non-trivial compositional dependencies within multiphase systems lies in incorporating prior domain knowledge or physical constraints, acquired by a complementary method, into the acquisition function. For example, incorporating phase boundary information, as demonstrated in the *closed-loop autonomous system for materials exploration and optimization* (CAMEO) framework, allows for targeted exploration near regions of significant property changes.[379]

**VI. Integration into broader workflows**

Materials are inherently complex, multiscale, and multiphysics systems. Their hierarchical organization means that understanding macroscopic behavior often requires uncovering multiple microscopic properties and intricate interdependencies between them.[380] This makes SPM, capable of probing a wide range of local properties, an extremely useful technique for material characterization. However, achieving a comprehensive understanding of material properties often necessitates the use of multiple characterization techniques. This necessity becomes even more pronounced in the case of combinatorial libraries, where the goal often lies in uncovering highly convoluted dependencies across compositional spaces. These considerations bring us to the point, where the development of integrated frameworks that combine multiple complementary techniques for material characterization and optimization is pivotal for further acceleration of material design. For instance, combining ESM with techniques like Time-of-Flight Secondary Ion Mass Spectrometry (ToF-SIMS) in an automated manner enables the exploration of relationships between local electrochemical behavior and material properties, such as ion diffusion coefficients and bias-dependent reaction rates. Such integration may also facilitate the identification of key microstructural features and help in defining their role in ionic and electronic transport. This provides critical insights for optimizing both material composition and processing conditions. The extension of these local methods by the macroscopic data like impedance measurements (IS) should allow for bridging localized and bulk material properties by the development of predictive models that can be used to guide automated exploration or automatization. Importantly, in this case, SPM plays the role of a low fidelity but very fast measurement enabling the rapid exploration of electrochemical responses as a function of composition across the composition-spread library. IS characterization is a middle fidelity but more time-consuming approach. The multifidelity frameworks should allow for accounting of the individual costs associated with each method, thereby reducing the overall expenses while enhancing the efficiency of the optimization process. Similarly, integrating SPM techniques with chemical mapping methods like Energy-Dispersive X-ray Spectroscopy (EDS) or crystallographic mapping via Electron Backscatter Diffraction (EBSD), or Raman Spectroscopy enables the correlation of functional properties with structural features such as grain boundaries and phase interfaces.

An automated experiment within a single modality for optimizing material properties across a combinatorial space is typically structured as an iterative process. In each iteration, an AI agent, such as the GP BO model, is employed to uncover composition-property relationships and predict the property values across the explored compositional space. Guided by these predictions, the agent selects the next composition (or location within the combinatorial library) to investigate aiming to achieve the predefined target with the fewest possible steps. Model ability to uncover structure-property relationships significantly reduces the number of observations needed for property optimization or exploration. In this scheme, characterizing materials using multiple modalities involves employing multiple AI agents, each independently guiding the optimization process for its respective modality. Although this experimental design enables faster material design compared to grid search methods, it offers limited opportunities for knowledge sharing between different characterization techniques used in material characterization. The knowledge gained from one method can only be incorporated as priors for another necessitating a sequential approach to combining them in a single automated experiment, as demonstrated in the CAMEO framework (Figure 15a).[379]

### VI.a. Co-orchestration of multiple tools

The next step toward enhanced optimization is integrating multiple modalities within a single experiment for the *simultaneous* exploration of the compositional space. This shift introduces a fundamentally new vision for the automated material characterization process, where characterization is no longer considered as a series of independent or almost independent automated experiments. Instead, it becomes a unified workflow that combines multiple modalities, collectively driving the entire exploration process toward predefined objectives. Thus, while multiple AI agents guide each modality independently in automated experiment based on single-modality optimizations, a single unified AI model is required to govern the entire experimental process for such a multimodal automated experiment. This model *orchestrates* the exploration across modalities, ensuring an efficient progression toward the experimental objectives. The core concept of multimodal methods is to uncover not only property-composition relationships but also the complex interdependencies between different modalities in an automated manner, enabling more accurate predictions of target functionalities.

The first proposed workflow for integrating different modalities into a single automated experiment with real-time data sharing was the MULTITASK framework.[381] By conceptualizing the automated lab as a modular structure, the authors illustrate how real-time knowledge sharing among different modalities can enhance and accelerate the optimization of target material properties across a compositional space.

Physical insights obtained from various microscopic techniques are often represented as high-dimensional datasets, such as spectra or images, rather than simple scalar values. In such cases, directly implementing BO necessitates the construction of multi-output surrogate models, which poses significant scalability challenges. A common approach to address this issue is scalarization, when meaningful scalar values are extracted from the initial high-dimensional data. For instance, this approach is exemplified in the MULTITASK framework, where Raman measurements are utilized to encode phase information, and the effective piezoelectric coefficient is extracted from PFM measurements. While the scalarization approach is robust and widely adopted, it presents several inherent limitations in the context of data sharing in multi-modal experiments.

Firstly, raw measurements often contain diverse insights into system behavior, and extracting only a single scalar value limits the potential to leverage other valuable information for accelerating complementary modalities. For example, typical PFM local ferroelectric hysteresis loops provide parameters such as the effective work of switching, bias, threshold (coercive) voltage, remanent polarization etc. A standard PFM scan contains a wealth of diverse information, ranging from roughness and topographical features to domain shapes, sizes, and their distributions. Scalarization, however, requires direct, expert-driven interpretation to extract specific parameters, often making it a complex and non-trivial task. In some cases, the absence of prior knowledge may hinder the identification of meaningful parameters, and focusing on a single parameter inevitably discards the remaining valuable information embedded in the dataset. This highlights the importance of developing approaches capable of leveraging all the information encoded in raw data.

One potential approach for unsupervised integration of multiple methods into a single automated workflow for exploring combinatorial libraries or guiding upstream synthesis is multimodal co-orchestration (Figure 15c).[382] This approach combines Variational Autoencoders (VAEs) for dimensionality reduction with Multitask Gaussian Processes (MTGP). A VAE is a neural network designed to map high-dimensional raw data into a low-dimensional latent space, effectively capturing the variational factors across the training dataset.[383] When variations in the acquired data are expected to be determined by compositional changes, VAE effectively encodes these relationships, extracting complex, convoluted features that are often beyond human interpretability. The interdependence of different properties measured by various techniques results in correlations between the compositional dependencies of their latent variables. Notably, each modality is independently encoded by its respective VAE. The MTGP captures these correlations and leverages them to enhance the prediction of the compositional dependencies of the latent variables for each modality. These predicted latent variables can then be decoded back into their high-dimensional representations, such as the original spectra or images, providing a comprehensive understanding of the system. Using MTGP, multimodal exploration can be integrated into a single automated experiment, where, at each iteration, the AI agent determines both the modality and the location (or composition) to be explored to maximize the value of each observation. This approach enables the design of highly time- and cost-efficient material exploration workflows by seamlessly combining different measurement techniques into a unified, autonomous process.

### VI.b. Theory co-navigation and digital twins

There is a growing interest within the scientific community in developing high-fidelity digital twins (DTs) for materials.[380] A DT is a detailed and dynamic digital representation of a specific material system, designed to replicate both its structural characteristics and functional responses. Typically, a DT comprises three key components: (1) a physical model that encapsulates the underlying principles governing the material behavior, (2) experimental data that inform and validate the physical model, and (3) a dynamic update protocol to amend the model with respect to newly acquired data.[384] These components collectively enable the DT to function as a "living" model, continuously evolving to capture the material system development and response to changing conditions.

One of the key challenges in developing a DT is establishing an effective feedback loop for real-time model tuning based on acquired data. A promising solution has been proposed in 2001 by M. C. Kennedy and A. O'Hagan for refining computer models.[385] This approach leverages BO to adjust the hyperparameters of theoretical models, progressively reducing epistemic uncertainty and aligning predictions with experimental observations. As the system evolves, these adjustments are informed by experimental results, leading to further refinements. In materials science, this iterative process is called *co-navigation* and enables the gradual adaptation of the physical model to the behavior of the actual material, ultimately transforming the theoretical framework into a fully realized digital twin (Figure 15b).[386]

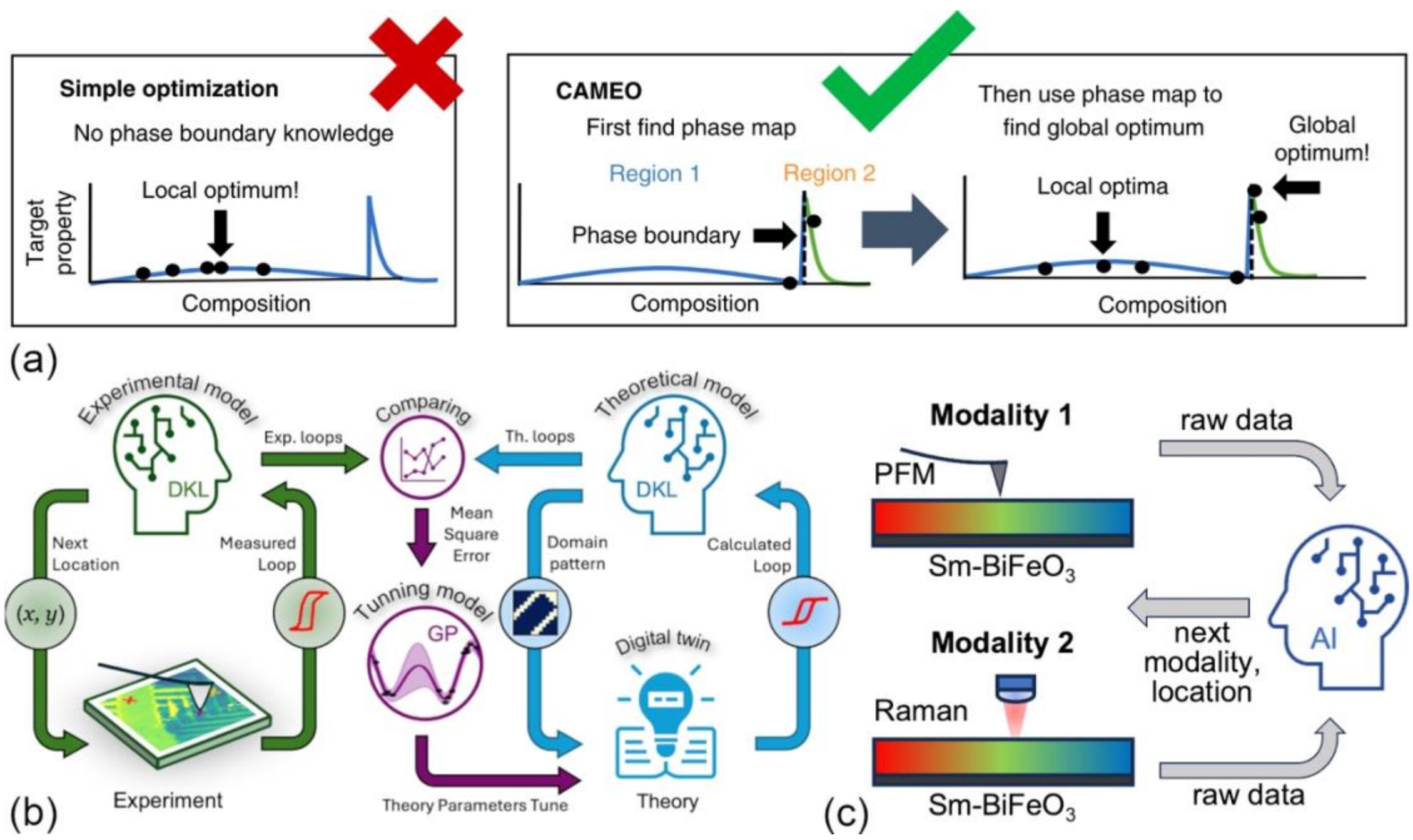

**Figure 15.** Automated workflow schemes. (a) Comparing simple optimization vs. CAMEO framework for automated combinatorial search. Reprinted with permission from [379] by A. G Kusne, et al., licensed under CC BY 4.0. (b) Bayesian co-navigation workflow for dynamic update of the theoretical model based on the experimental results. Reprinted with permission from.[386] Copyright (2024) American Chemical Society. (c) Multimodal Co-orchestration workflow for combinatorial libraries characterization. Reprinted from [382] by B. Slautin, et al., licensed under CC BY-NC 3.0.

The broad spectrum of properties that SPM can probe makes it an ideal candidate for integration into co-navigational loops. A notable demonstration of this approach was recently achieved in the context of local ferroelectric property exploration.[386] In this framework, SPM was seamlessly integrated into a co-navigational system, where its measurements were used to iteratively update the model parameters, reducing uncertainty and enhancing the accuracy of property predictions. This advancement highlights the SPM potential not only as a characterization tool but also as a dynamic component in feedback-driven workflows accelerating the discovery and optimization of material functionalities.

**Summary**

Scanning Probe Microscopy (SPM) emerged over 35 years revolutionizing experimental condensed matter physics by enabling unprecedented insights into structural properties at the nano- and mesoscale levels. The diversity of the measured properties and wide range of measurement conditions and environments make SPM indispensable for studying multiple functional phenomena. The advancement of SPM methods has coincided with progress in electronics and computer technologies driving a continuous expansion in the technique capabilities over the past decades.

The rise of self-driving labs and combinatorial materials synthesis poses a new challenge for the fast characterization of synthesized materials. In this review, we argue that SPM techniques are uniquely suited to meet this demand offering low-cost and powerful approaches to close the loop in high-throughput material design. Advances in ML approaches and AI-driven automated experimentation enable bridging the microscopic insights of SPM with traditional macroscopic characterization techniques and make it possible to overcome many of the traditional SPM limitations. We anticipate that the synergy between advanced AI approaches, combinatorial synthesis, and SPM will transform the fields of material design and optimization in the coming years, enabling closed-loop automated synthesis and characterization workflows.

**Disclosure**

Certain equipment, instruments, software, methods or materials are identified in this paper in order to specify the experimental procedures adequately. Such identification is not intended to imply recommendation or endorsement of any product or service by NIST, nor is it intended to imply that the materials or equipment identified are necessarily the best available for the purpose.

**Data availability**

Data sharing not applicable – no new data generated.

**Acknowledgements**

SVK and PDR acknowledge the support by the National Science Foundation Materials Research Science and Engineering Center program through the UT Knoxville Center for Advanced Materials and Manufacturing (DMR-2309083). AD is grateful to German Academic Exchange Service (Deutscher Akademischer Austauschdienst: DAAD-PRIME 2023-24) postdoctoral funding program. YL acknowledge the support the Center for Nanophase Materials Sciences (CNMS), which is a US Department of Energy, Office of Science User Facility at Oak Ridge National Laboratory. S.H. acknowledges the support from the National Research Foundation of Korea (NRF) grant funded by the Korea government (MSIT) (RS-2023-00247245). M.A. and S.L.S. acknowledge support from National Science Foundation (NSF), Award Number No. 2043205 and Alfred P. Sloan Foundation (Award No. FG-2022-18275). S.L.S. acknowledges partial support from the Center for Materials Processing (CMP) at the University of Tennessee, Knoxville. YK acknowledges the support by the National Research Foundation of Korea (NRF) grant funded by the Korea government (MSIT)(RS-2024-00346128). K.A.B. acknowledges support from the Army Research Office through W911NF2410093.